\documentclass[prd,aps,floats,floatfix]{revtex4}
\usepackage{amsmath}
\usepackage{amsfonts}
\usepackage{graphicx}
\begin{document}
\newcommand{\vect}[1]{\overrightarrow{#1}}
\newcommand{\smbox}[1]{\mbox{\scriptsize #1}}
\newcommand{\tanbox}[1]{\mbox{\tiny #1}}
\newcommand{\vev}[1]{\langle #1 \rangle}
\newcommand{\Tr}[1]{\mbox{Tr}\left[#1\right]}
\newcommand{\cosb}{c_{\beta}}
\newcommand{\sinb}{s_{\beta}}
\newcommand{\tanb}{t_{\beta}}
\newcommand{\picwidth}{3.4in}

\preprint{MSU-HEP-080508}
\title{Triplet Extended Supersymmetric standard model}

\author{Stefano Di Chiara}
\email[]{dichiara@msu.edu}
\author{Ken Hsieh}
\email[]{kenhsieh@pa.msu.edu}
\affiliation{Department of Physics,
Michigan State University, East Lansing, Michigan 48824, USA}
\date{\today}

\begin{abstract}
We revisit an extension of the MSSM
by adding a hypercharge-neutral, $SU$(2)-triplet chiral
superfield.
Similar to the NMSSM, the triplet gives an additional contribution
to the quartic coupling in the Higgs potential, and the mass of
the lightest $CP$-even Higgs boson can be greater than $M_Z$ at
tree-level.
In addition to discussing the perturbativity, fine-tuning, and decoupling
issues of this model,
we compute the dominant 1-loop corrections to the
mass of the lightest $CP$-even Higgs boson
from the triplet sector.
When the Higgs-Higgs-Triplet coupling in the superpotential
is comparable to the top Yukawa coupling, we find that the Higgs
mass can be as heavy as 140 GeV even without
the traditional contributions from the top--s-top
sector, and at the same time consistent with
the precision electroweak constraints.
At the expense of having Landau poles before
the GUT scale, this opens up a previously forbidden
region in the MSSM parameter space where both
s-tops are light.
In addition to having relatively small fine-tuning
(about one part in 30),
this leads to a gluo-philic
Higgs boson whose production via gluon-gluon fusion
at the LHC
can be twice as large as the SM prediction.
\end{abstract}
\maketitle

\section{Introduction}
The electroweak sector of the standard model (SM)
predicts new physics at sub-TeV scales
to unitarize $WW$-scattering.
With a Higgs boson, the hierarchy problem suggests
additional new physics near the TeV scale
to stabilize the electroweak scale,
and the minimal supersymmetric standard model (MSSM)
is one of the leading candidates of such new physics.
For reviews of the MSSM, see, for examples,
Drees \cite{Drees:1996ca}, Martin \cite{Martin:1997ns},
Dine \cite{Dine:1996ui}, and Peskin \cite{Peskin:2008nw}.

In the MSSM, the mass of the lightest $CP$-even
Higgs boson is bounded at tree-level by $M_Z$,
because the tree-level quartic couplings are
parameterized by gauge couplings,
and such a light Higgs boson is ruled out
by the CERN LEP searches of the SM Higgs boson
\cite{Acciarri:2001si}\cite{Heister:2001kr}\cite{Abbiendi:2002yk}\cite{Abdallah:2003ip}\cite{Barate:2003sz}
that impose
\begin{align}
m_h^{\smbox{SM}}> 114.4\ \mbox{GeV}.
\end{align}
At one-loop level, however, there can be large radiative
corrections due to heavy scalar tops ($\widetilde{Q}_3$ and
$\widetilde{\overline{U}}_3$, superpartners of the top-quark)
and/or a large coupling of the trilinear interaction
$\widetilde{Q}_3H_u\widetilde{\overline{U}}_3$
\cite{Okada:1990vk}\cite{Haber:1990aw}\cite{Haber:1993an}\cite{Haber:1996fp}\cite{Heinemeyer:1998kz}\cite{Heinemeyer:1999be}\cite{Heinemeyer:1999zf}\cite{Espinosa:1999zm}\cite{Carena:2000dp}\cite{Espinosa:2000df}.
While such radiative corrections
can be large enough to satisfy the LEP
bounds, they also contribute
to the quadratic term of the Higgs potential,
leading to the ``little hierarchy problem''.
The MSSM also suffers from a $\mu$-problem
in that its lone dimensionful SUSY-invariant
parameter, $\mu$,
is phenomenologically required to be of
order 100 GeV, while its natural scale
can in principle be much larger.

The next-to-minimal supersymmetric standard model
(NMSSM) solves the $\mu$-problem and alleviates
the little hierarchy problem \cite{BasteroGil:2000bw}
by
extending the MSSM with a
singlet chiral superfield $S$.
For reviews of the NMSSM, see Balazs et al.~\cite{Balazs:2007pf}
and references therein.
The Higgs couplings with $S$ lead
to additional contributions
to the quartic couplings in the Higgs potential,
while the $\mu$-term is dynamically generated
from the vacuum expectation value (vev) of the scalar
component of $S$.
With these additional contributions to the quartic
couplings, the mass of the lightest $CP$-even Higgs
boson may be larger than $M_Z$ at tree-level,
and the NMSSM can satisfy the LEP bounds
on the Higgs mass with lighter s-tops compared to the MSSM
\cite{Pandita:1993tg}\cite{Pandita:1993hx}\cite{Elliott:1993bs}\cite{Codoban:2002rh}\cite{Miller:2003ay}\cite{Ellwanger:2005fh}\cite{Ellwanger:2006rm}.

In this paper, we extend the MSSM with a
hypercharge-neutral, $SU(2)$-triplet
chiral superfield $T$ and name the model
triplet-extended supersymmetric standard model (TESSM).
Extensions of this type have been studied extensively by Espinosa
and Quiros \cite{Espinosa:1991wt}\cite{Espinosa:1991gr},
Felix-Beltran \cite{FelixBeltran:2002tb}, Setzer and Spinner \cite{Setzer:2006sf},
and Diaz-Cruz et al.~\cite{DiazCruz:2007tf}.
While this model does not solve the $\mu$-problem,
it is an interesting alternative to the NMSSM,
as an economical extension of the MSSM, because
it can also achieve a mass of the lightest $CP$-even Higgs boson
that is larger than $M_Z$ at tree level.
Furthermore, compared to the MSSM and the NMSSM,
we expect there to be more radiative corrections to
the mass of the lightest $CP$-even Higgs boson due to the additional
states in the triplet.
To the extent that these triplet-induced radiative
corrections are significant, we may further alleviate the little hierarchy
problem.

Unfortunately, in both the NMSSM and the TESSM,
the respective singlet-induced and triplet-induced radiative corrections
are typically small when we demand perturbativity at the scale of grand unified theory (GUT)
near $10^{16}$ GeV.
This is because perturbativity at the GUT scale
imposes the bound $\lambda \lesssim 0.7$ at the weak scale,
where $\lambda$ is respectively the Higgs-singlet-Higgs
and the Higgs-triplet-Higgs coupling in the superpotential of
the NMSSM and TESSM.
In both models, while the tree-level mass of the lightest $CP$-even Higgs boson
can be as large as 100 GeV,
the $\mathcal{O}(\lambda^4)$ radiative corrections
are not large enough to lift the Higgs mass over the LEP bounds.
On the other hand, in the TESSM,
when we have $\lambda\sim 0.9$ (so that $\lambda$ is comparable with the
top Yukawa coupling)at the weak scale,
we find the tree-level mass of the lightest $CP$-even Higgs boson
to be close to the LEP bound and the $\mathcal{O}(\lambda^4)$ radiative
corrections alone can easily lift the Higgs mass over the LEP bound even
with small SUSY-breaking in the triplet sector.
As the small SUSY-breaking in the triplet sector translate into
small fine-tuning, we can solve the little hierarchy problem at
the expense of giving up perturbativity at the GUT scale.

Without demanding perturbativity at the GUT scale,
we also expect the NMSSM to be a solution to the little hierarchy problem,
with the mass of the lightest $CP$-even Higgs boson that
satisfies the LEP bounds without significant contribution from
the top--s-top sector.
However, as an alternative to the NMSSM and a reasonably economical extension of the MSSM,
the TESSM and its phenomenology are interesting in their own right.
For example, as we show in this paper, the MSSM limit of the TESSM
is achieved with $M_T\rightarrow \infty$, where $M_T$ is the SUSY-invariant
mass of the triplet, keeping $\lambda$ fixed, whereas in the NMSSM one requires
$\lambda\rightarrow 0$ to achieve the MSSM limit.
As another example, even though
the sub-TeV, electrically-neutral component of the triplet acquires a vev, we can still
satisfy the precision
electroweak constraints without the extreme fine-tuning noted in
the triplet-extended SM \cite{SekharChivukula:2007gi}.
Moreover, there may be other considerations that motivate extending
the MSSM by a triplet instead of a singlet.
For example, in obtaining neutrino masses through the Type-II \cite{Mohapatra:1980yp} and Type-III
seesaw mechanisms, the SM is commonly extended with Higgs triplets.
Though the Higgs triplets may have nonzero hypercharge, hypercharge-neutral
triplets are often present when the models are supersymmetrized
and embedded in a unified gauge group \cite{Setzer:2006sf}\cite{Mohapatra:2007js}\cite{Mohapatra:2008gz}.

We organize our paper as follows.
In Sec.~\ref{sec:TESSM}, we lay out the superpotential
and the Lagrangian of the TESSM, compare it to the NMSSM,
and discuss constraints on
its parameter space from electroweak precision tests and the requirement
of perturbativity at the GUT scale.
In Sec.~\ref{sec:HiggsMass}, we numerically
evaluate the mass of the lightest, $CP$-even Higgs boson
to one-loop, and show that we can satisfy the LEP2 bounds
without the contributions from the top--s-top sector when $\lambda$ is
comparable with the top Yukawa coupling.
We also discuss the gluon-gluon fusion production
and diphoton decay of the lightest, $CP$-even Higgs boson
in Sec.~\ref{sec:HiggsMass}.
Our discussions of the gluon-gluon fusion production
rely only on the existence of light s-tops and the
minimal color sector of the MSSM, and are therefore applicable to
any extensions of the MSSM that solves the little hierarchy problem
without invoking additional colored states.
In Sec.~\ref{sec:FineTuning}, we estimate two sources of fine-tuning
in this model, and find that we can achieve a small fine-tuning of about one part
in 30 in the Higgs sector.
Finally, we conclude with Sec.~\ref{sec:Conclusion} that summarizes
our results.

\section{Triplet-Extended Supersymmetric Standard Model}
\label{sec:TESSM}
\subsection{The Model}
We extend the MSSM with a hypercharge-neutral, $SU(2)$-triplet
$T\equiv\tfrac{1}{2}\sigma^AT^A$ with the superpotential
\begin{align}
W_{\smbox{TESSM}}= \mu H_d H_u + M_T \mbox{Tr}(T\,T) +2\lambda H_d T
H_u + \alpha_T \mbox{Tr}(T) +W_{\smbox{Yukawa}}, \label{eq:MODEL}
\end{align}
where $H_{u,d}$ are the Higgs doublets of the MSSM, $\alpha_T$ is
a Lagrange-multiplier determined from the potential, and
$W_{\smbox{Yukawa}}$ is the MSSM superpotential sans the
$\mu$-term
\begin{align}
W_{\smbox{Yukawa}}= y_t Q H_u U^c + y_b Q H_d D^c + y_{\tau} L H_d E^c. \label{eq:YUKAWA}
\end{align}
Note that, since $T$ is a chiral superfield, its scalar component
necessarily contains a \emph{complex} $SU(2)$-triplet, whereas
in non-SUSY extensions of the SM
\cite{Blank:1997qa}\cite{Chen:2005jx}\cite{Chen:2006pb}\cite{Chankowski:2006hs}\cite{Chankowski:2007mf}\cite{SekharChivukula:2007gi}, we can extend the SM with a real $SU(2)$-triplet.
In components, we have the fields
\begin{align}
H_u=\begin{pmatrix}H_u^+ \\ H_u^0\end{pmatrix}, \quad
H_d=\begin{pmatrix}H_d^0 \\ H_d^-\end{pmatrix}, \quad
T=\frac{1}{2}
\begin{pmatrix}T^0 & \sqrt{2}T^{+}\\ \sqrt{2}T^{-} & -T^0\end{pmatrix},
\end{align}
and the superpotential (sans the SM Yukawa couplings)
\begin{align}
W_{\smbox{TESSM}}&\supset \mu(H_u^+ H_d^- - H_u^0H_d^0)
+\frac{M_T}{2}(T^0T^0+2T^{+}T^{-}) \nonumber\\
&\quad+
\lambda(H_d^0 T^0 H_u^0+H_d^- T^0 H_u^+)
+\sqrt{2}\lambda(H_d^{-} T^{+} H_u^{0}-H_d^{0} T^{-} H_u^{+}).
\end{align}
The factor of 2 in front of $\lambda$ in Eq.~\ref{eq:MODEL}
gives us
a coefficient of unity for the term
\begin{align}
W_{\smbox{TESSM}}\supset \lambda H_d^0 T^0 H_u^0,
\end{align}
as with the case of the NMSSM when $T^0$ is replaced
by a singlet $S$, and facilitates direct
comparisons between TESSM and NMSSM.

We can achieve gauge coupling unification at $M_{\smbox{GUT}}$
by including additional chiral superfields with
quantum numbers
\begin{align}
D\sim (\mathbf{1},\mathbf{2})_{\frac{1}{2}},
\quad
\overline{D}\sim (\mathbf{1},\mathbf{2})_{-\frac{1}{2}},
\quad
G\sim (\mathbf{8},\mathbf{1})_{0},
\label{eq:AddedX}
\end{align}
where the first and second entries inside the parenthesis
denote, respectively, the representations
under the color $SU(3)_c$ and weak $SU(2)_w$
gauge groups, and the subscripts
denote the charge under hypercharge $U(1)_Y$ gauge group.
This added content can have both
SUSY-invariant and SUSY-breaking masses sufficiently large (say 2 TeV's) so that
they decouple from the electroweak scale physics,
while still allowing for gauge coupling unification.
The added matter content (triplet plus those in Eq.~\ref{eq:AddedX})
does not constitute a complete multiplet of SU(5),
but can form a complete multiplet of
trinification group
$SU(3)^3\times Z_3$ \cite{Babu:1985gi}\cite{Willenbrock:2003ca}\cite{Sayre:2006ma}.

In addition to the MSSM soft SUSY-breaking parameters, we also
have soft terms involving $T$
\begin{align}
-\Delta\mathcal{L} = 2m_T^2\mbox{Tr}(T^{\dag}T) +B_T
(\mbox{Tr}(T\,T)+\mbox{h.c.}) +2\lambda A_{\lambda} (H_d T
H_u+\mbox{h.c.}).
\end{align}

\subsection{Comparison to the NMSSM}
\subsubsection{Perturbativity}
For simplicity, we assume that all couplings  and
masses in the superpotential are real.
The tree-level potential involving the $U(1)_{\smbox{em}}$-neutral Higgs doublets and triplet
is then
\begin{align}
V_{\smbox{TESSM}}=V_{H}+V_{T}+V_{\smbox{mix}},
\label{eq:TreePotential}
\end{align}
where
\begin{align}
V_{H}&= (\mu^2+m^2_{H_u})|H_u^0|^2+(\mu^2+m^2_{H_d})|H_d^0|^2
-B_{\mu}(H_u^0H_d^0+\mbox{c.c.})\nonumber\\
&+
\frac{1}{8}(g_2^2+g_1^2)(|H_u^0|^2-|H_d^0|^2)^2
+\lambda^2|H_u^0|^2|H_d^0|^2,
\label{eq:VH}
\\
V_{T}&= (M_T^2+m^2_T)|T^0|^2+\frac{B_T}{2}(T^0T^0+\mbox{c.c.}),
\label{eq:VT}
\\
V_{\smbox{mix}}&=
\lambda^2|T^0|^2(|H^0_u|^2+|H^0_d|^2)\nonumber\\
&+
\lambda M_T(H^0_d T^{0\ast} H^0_u+\mbox{c.c.})
+\lambda A_{\lambda}(H^0_d T^{0}H^0_u+\mbox{c.c.})
\nonumber\\
&-\lambda\mu(H^{0\ast}_u T^0 H_u^0+H^{0\ast}_d T^0 H_d^0
+\mbox{c.c.}).
\label{eq:VMIX}
\end{align}

Compared with the Higgs potential in the MSSM,
we have
an enhancement in the quartic coupling of the form
\begin{align}
V\supset \lambda^2 |H_u|^2 |H_d|^2,
\label{eq:MainQ}
\end{align}
and this in principle allows for a tree-level mass
eigenvalue larger than $M_Z$ after
electroweak symmetry breaking (EWSB).
This is similar to the case in the NMSSM,
where such quartic couplings are also generated
from a superpotential of the form
\begin{align}
W_{\smbox{NMSSM}}=\lambda S H_d H_u+\frac{\kappa}{3}S^3.
\end{align}
As with the NMSSM, where
the requirement of perturbativity
at the GUT scale limits $\lambda \lesssim 0.7$
at the TeV-scale,
the TESSM  also has a bound $\lambda\lesssim 0.7$
at the TeV-scale while still preserving perturbativity
at the GUT scale.
Though the bounds are similar, the details of obtaining
such bounds are different and may be important
for further model-building where perturbativity
at the GUT scale is imposed.
We elaborate briefly on some key differences between
the two models from the relevant
renormalization group equations (RGEs)
given by
\begin{align}
\beta_{\lambda}^{\smbox{TESSM}}&=
\frac{\lambda}{16\pi^2}
\left(8\lambda^2+3 y_t^2-9g_2^2-g_1^2\right),
\\
\beta_{\lambda}^{\smbox{NMSSM}}&=
\frac{\lambda}{16\pi^2}
\left(4\lambda^2+2\kappa^2+3 y_t^2-3g_2^2-g_1^2\right),
\\
\beta_{\kappa}^{\smbox{NMSSM}}&=
\frac{\kappa}{16\pi^2}
\left(6\lambda^2+6\kappa^2\right),
\end{align}
and note the following points:
\begin{itemize}
\item
In the NMSSM, there are two possible Landau poles in
$\lambda$ and $\kappa$.
The RGE of $\kappa$
is such that $\kappa$ always increases when
evolving to higher energies, and $\kappa$ feeds into the
evolution of $\lambda$.
In the TESSM, there is no such contribution
because
Tr$(T^3)=0$, but there are now additional
contributions to
the $\lambda^3$ coefficient in $\beta_{\lambda}$
in the TESSM.
\item
In the TESSM, $\beta_{\lambda}$
has a larger coefficient for the negative contribution
of the form $\lambda\ g_2^2$
because $T$ is charged under $SU(2)$.
As the coupling $g_2$ is non-asymptotically-free in the TESSM (also
in the NMSSM), this gives a stronger suppression
at higher energies and could potentially delay the appearance of the $\lambda$ Landau pole.
\item
The coupling $g_2$ also flows to larger values
in the TESSM than in the NMSSM because of the added matter
content.  This again gives a suppression at higher energies
and may delay the $\lambda$ Landau pole.
(This can be achieved in the NMSSM with added matter content,
for example, in the NMSSM with gauge-mediated SUSY-breaking.)
\end{itemize}

Thorough studies on the upper bounds of $\lambda$ in TESSM
and its extensions would require examining fixed points from the RGEs,
and we leave these investigations for the future.
For our work, it suffices to note that perturbativity at the GUT scale
imposes $\lambda\lesssim 0.7$ at the weak scale, so that
$\lambda$ is of similar strength to the weak gauge coupling.
As such, while the tree-level mass of the lightest $CP$-even Higgs boson
can be 100 GeV (as we later show), we expect the $\mathcal{O}(\lambda^4)$ radiative corrections to the mass
of the lightest $CP$-even Higgs boson to be insufficient to lift the Higgs mass above
the LEP bounds.
However, motivated by solving the hierarchy problem, we take the view point that
the Landau pole we encounter at a higher scale (around $10^{10}$ GeV) is rescued by some other new physics
and analyze
the Higgs spectra and the phenomenology for larger values of $\lambda$.
We take values of $\lambda$ comparable to the top Yukawa coupling, so that the
TeV scale physics is still perturbative.
As $\lambda$ is now near unity at the TeV-scale in the TESSM
and we expect there to be more $\mathcal{O}(\lambda^4)$ radiative corrections to the mass
of the lightest $CP$-even Higgs boson compared to the NMSSM, it is worthwhile to investigate
these radiative corrections in detail.

In extensions of the NMSSM, such as fat Higgs models
\cite{Batra:2003nj}\cite{Harnik:2003rs}\cite{Chang:2004db}\cite{Delgado:2005fq},
$\lambda$ can achieve much larger values and give rise to a very large
mass for the
lightest $CP$-even Higgs boson.
The model-building techniques of fat Higgs models can also be
applied to the TESSM, but in this work we focus on
the TESSM as a simple extension of the MSSM and an alternative
to the NMSSM without imposing the constraint of perturbativity at the GUT
scale.

\subsubsection{The MSSM limit}
In the NMSSM, the $\mu$-term and the masses of the singlet(ino)
are related by
\begin{align}
\mu \sim \lambda^{\smbox{NMSSM}}\langle S\rangle,\quad
M_S \sim \kappa^{\smbox{NMSSM}}\langle S\rangle,
\end{align}
and the MSSM limit is $M_S\rightarrow \infty$ while
keeping $\mu$ fixed.
Keeping $\kappa$ perturbative in the MSSM limit then
gives $\lambda\rightarrow 0$.  As $\lambda$ is the only
coupling between the singlet and the MSSM sector, the
MSSM limit of $\lambda\rightarrow 0$ with fixed $\mu$
decouples the singlet.

\begin{figure}[t]
\begin{center}
\includegraphics[width=2in]{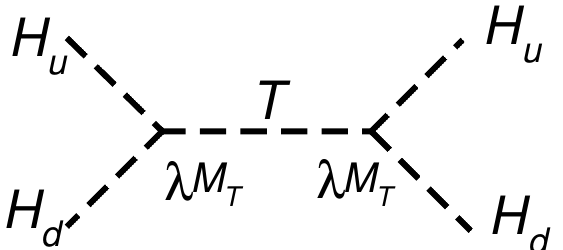}
\caption{An example of Feynman diagrams that
give the contributions which cancel
$\lambda$ contributions in the Higgs potential
when the triplet field, $T$, decouples.
}
\label{fig:effop}
\end{center}
\end{figure}

In the TESSM, the MSSM limit is achieved with
$M_T\rightarrow\infty$,
holding the values of all other masses and couplings at the
weak scale fixed, and in particular we do not need
$\lambda\rightarrow 0$ to achieve the MSSM limit.
The decoupling of the additional contribution
to the Higgs quartic coupling
in Eq.~\ref{eq:MainQ} is accomplished by the effective operator
obtained by integrating out the heavy triplet
fields when $M_T\gg M_{Z}$.
Setting $B_T=0$ for simplicity,
the equation of motion for $T^0$, among other
terms, has contributions of the form
\begin{align}
T^0=-\frac{\lambda}{M^2_T+m^2_T}
\left(
M_T H_d^0H_u^0
+A_{\lambda}H_d^{0\ast}H_u^{0\ast}-
\mu(H_u^{0\ast}H_u^0+H_d^{0\ast}H_d^0)
\right)+\cdots,
\end{align}
and this induces a contribution in the
effective Lagrangian
\begin{align}
-\Delta\mathcal{L}_{\smbox{eff}}=
-\lambda^2\frac{M_T^2+A_{\lambda}^2+2\mu^2}{M_T^2+m_T^2}|H_d^0|^2|H_u^0|^2
+\cdots,
\end{align}
that cancels the $\lambda^2$ contribution
to the quartic in the Higgs potential
when $M_T^2\gg A^2_{\lambda}, m^2_T,\mu^2$.
In terms of Feynman diagrams, this effective operator
arises from the diagrams such as the one shown in Fig.~\ref{fig:effop}
with the amplitude (in the limit of large $M_T^2\gg A^2_{\lambda}, m^2_T,\mu^2$)
\begin{align}
i\mathcal{A}=\lambda^2 \frac{M_T^2}{p^2-M_T^2},
\end{align}
where $p\sim M_Z$ is the scale of external
momenta of the Higgs bosons.
In the limit $M_T^2\gg p^2$, this gives
the canceling contribution to the Higgs potential,
and the resulting theory is the MSSM.

When we do not explicitly integrate out the heavy triplet sector,
the full mass matrices (involving both Higgs doublets and the triplet)
provide a seesaw-like mechanism in the limit of $M_T\rightarrow \infty$
that seesaws away any $\lambda$ dependence in the Higgs doublets sector.
We will demonstrate this in the next section when we compute
the mass of the lightest $CP$-even Higgs boson.

\subsection{EWSB in TESSM}
As we are assuming real couplings and masses for simplicity,
there is no mixing between the real and imaginary components
of the complex scalar fields $H^0_{u,d}$ and $T^0$ and
it is
convenient to separate them into real and imaginary
parts
\begin{align}
H^0_u&=\frac{1}{\sqrt{2}}(a_u + \imath b_u)=
\frac{1}{\sqrt{2}}(a^{\prime}_u + \imath b_u) +\frac{1}{\sqrt{2}}v_u,\\
H^0_d&=\frac{1}{\sqrt{2}}(a_d + \imath b_d)=
\frac{1}{\sqrt{2}}(a^{\prime}_d + \imath b_d) +\frac{1}{\sqrt{2}}v_d,\\
T^0  &=\frac{1}{\sqrt{2}}(a_t + \imath b_t)=
\frac{1}{\sqrt{2}}(a^{\prime}_t + \imath b_t) +\frac{1}{\sqrt{2}}v_t,
\end{align}
where we have also shifted the real components ($a_i$) to
the physical modes ($a^{\prime}_i$) by the
respective vacuum expectation values ($v_i$).
Prior to EWSB, all the vevs vanish
and the real components of the Higgses have the mass matrix (in
the basis $(a_u,a_d,a_t)$)
\begin{align}
\mathcal{M}^2_{a}=
\begin{pmatrix}
m_{H_u}^2+\mu^2 & -B_{\mu} & 0 \\
-B_{\mu} & m_{H_d}^2+\mu^2 & 0 \\
0 & 0 & M_T^2+m_T^2+B_T
\end{pmatrix}.
\label{eq:v0RealMatrix}
\end{align}
The corresponding mass matrix for the imaginary components can be
obtained from Eq.~\ref{eq:v0RealMatrix} by changing the signs of
$B_T$ and $B_{\mu}$ in the (1,2), (2,1), and (3,3) elements.

In the MSSM, the conditions for successful EWSB breaking are
that (i) the (top-left 2$\times$2) mass matrix in Eq.~\ref{eq:v0RealMatrix} has one
negative eigenvalue, and (ii) the potential is bounded from below
along the $D$-flat direction $H_u^0=H_d^0$.
In the TESSM model, the first condition gives us the same
condition as the MSSM
\begin{align}
B_{\mu}^2 > (m_{H_u}^2+\mu^2)(m_{H_d}^2+\mu^2)
\label{eq:EWConstrt},
\end{align}
while the second condition is automatically satisfied by the
presence of the quartic coupling $\lambda^2 |H_d^0|^2 |H_u^0|^2$.
However, the minimization conditions now demand
\begin{align}
v_t &= \frac{\sqrt{2}}{2}
(\lambda v^2)\
\frac{\mu-(A_{\lambda}+M_T)\cosb\sinb}
{M_T^2+m_T^2+B_T+\tfrac{\lambda^2}{2}v^2},
\label{eq:vtCond}
\\
m_{H_u}^2+\mu_{\smbox{eff}}^2&= \tanb^{-1}B_{\mu}+\frac{c_{2\beta}}{2}M_Z^2-
\frac{1}{2}\lambda^2\cosb^2v^2-\frac{\lambda
v_t}{\sqrt{2}}
(M_T+A_{\lambda})\tanb^{-1},
\label{eq:Min-MHU}\\
m_{H_d}^2+\mu_{\smbox{eff}}^2&= \tanb B_{\mu}-\frac{c_{2\beta}}{2}M_Z^2-
\frac{1}{2}\lambda^2\sinb^2v^2-\frac{\lambda
v_t}{\sqrt{2}} (M_T+A_{\lambda})\tanb
\label{eq:Min-MHD},
\end{align}
where we have defined
\begin{align}
\mu_{\smbox{eff}}&\equiv \mu - \frac{1}{\sqrt{2}}\lambda v_t,
\label{eq:mueff}\\
\tan\beta &\equiv \frac{v_u}{v_d},\quad v^2 \equiv v_u^2 + v_d^2,\\
v_u &= v \sin\beta,\quad v_d = v \cos\beta,
\end{align}
so that the gauge bosons receive masses of
\begin{align}
M_Z^2&=\frac{1}{4}(g_2^2+g_1^2)v^2,\\
M_W^2&=\frac{1}{4}g_2^2 v^2 + g_2^2 v_t^2,
\end{align}
where $g_2$ and $g_1$ are respectively the gauge couplings of the
$SU(2)$ and $U(1)_Y$ groups.
We have also abbreviated for convenience
the trigonometric functions
\begin{align}
s_{\beta}&\equiv\sin\beta,\quad c_{\beta}\equiv\cos\beta,\quad
t_{\beta}\equiv\tan\beta,\nonumber\\
s_{2\beta}&\equiv\sin 2\beta,\quad c_{2\beta}\equiv\cos 2\beta.
\end{align}

\subsection{Oblique Corrections}
While the condition of successful EWSB
in Eq.~\ref{eq:EWConstrt} gives a constraint
on the parameters, electroweak precision
tests offer a much more stringent constraint.
The induced vev $v_t$ contributes to the oblique parameter
$\alpha T$ because it contributes to the mass of the charged gauge bosons
$W^{\pm}$, but not to that of the neutral gauge boson $Z$.
We find the oblique contribution to be
\begin{align}
\alpha \Delta T&=\frac{\delta M_W^2}{M_W^2}=4\frac{v^2_t}{v^2}
\nonumber\\
&=\frac{\lambda^2\,v^2}{2}
\left(
\frac{2\mu-(A_{\lambda}+M_T)\sin 2\beta}
{M_T^2+m_T^2+B_T+\tfrac{\lambda^2}{2}v^2}
\right)^2.
\label{eq:ObliqueTFormula}
\end{align}
The oblique correction due to the triplet vanishes in
the limit of $M_T\rightarrow\infty$ holding
all other parameters fixed, as expected.
However, even if $M_T$ is of the same order of $\mu$ and $A_{\lambda}$,
$\Delta T$ can be small
due to a partial cancellation
between $\mu$ and $\sin 2\beta(A_{\lambda}+M_T)$.

\begin{figure}[t]
\begin{center}
\includegraphics[width=5in]{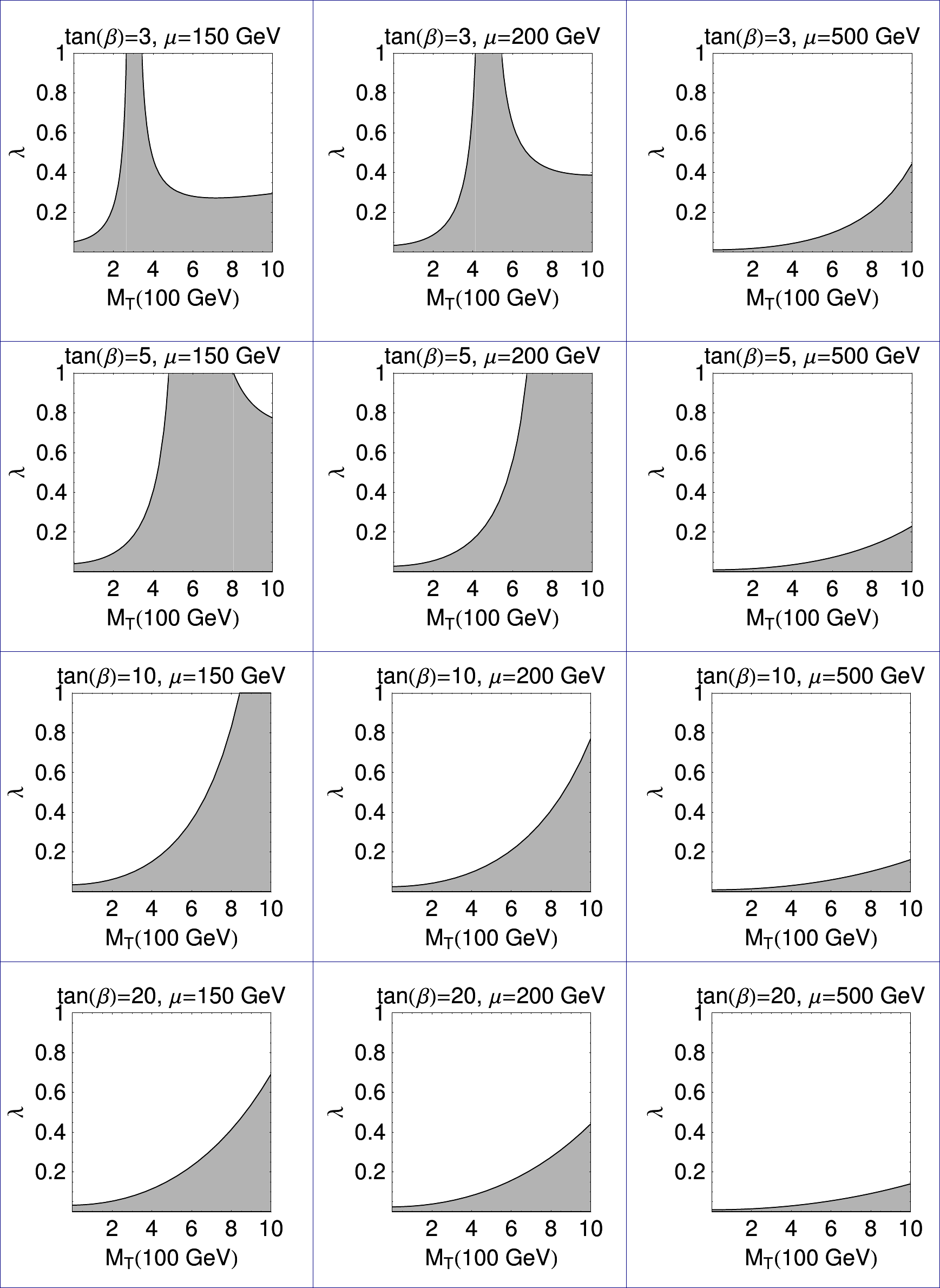}
\caption{
Regions allowed by $\Delta T$ (in gray) on $\lambda-M_T$ plane
for various values of $\tan\beta$ and $\mu$ as
indicated in each plot.
}
\label{fig:TPlot1}
\end{center}
\end{figure}

\begin{figure}[bt]
\begin{center}
\includegraphics[width=5in]{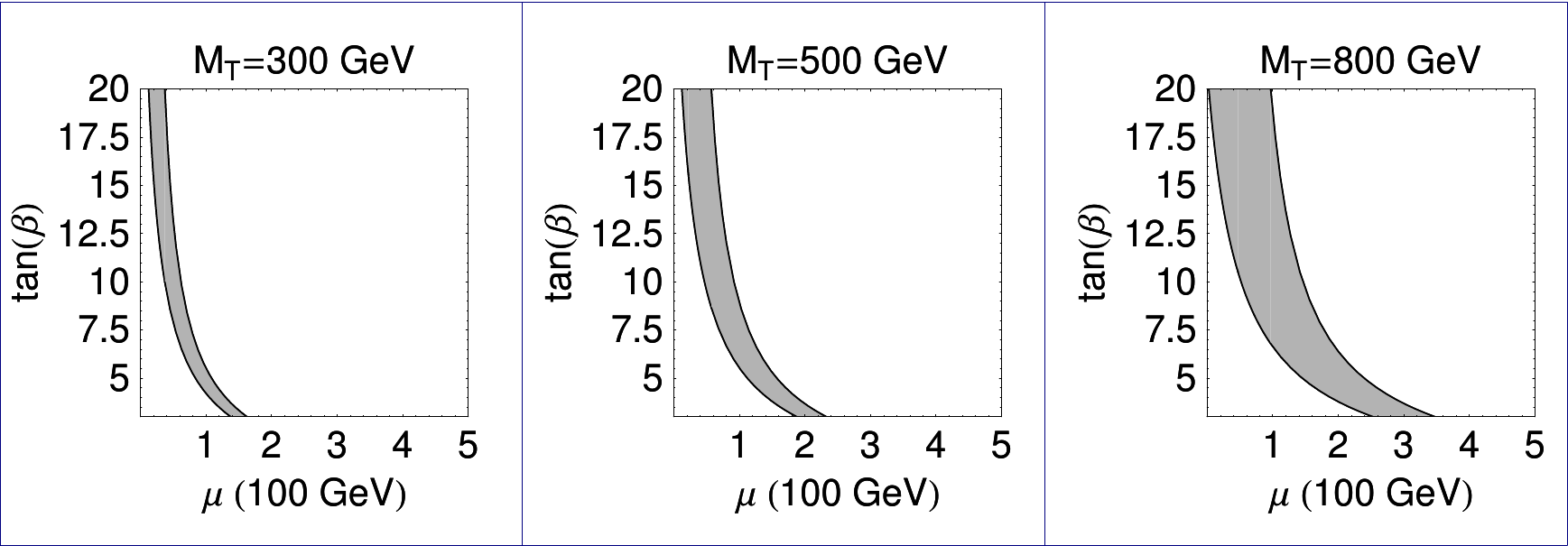}
\caption{
Regions allowed by $\Delta T$  (in gray) on $\tan\beta-\mu$ plane
for various values of $M_T$ as
indicated in each plot.
}
\label{fig:TPlot2}
\end{center}
\end{figure}

We impose the constraint that $|\Delta T|< 0.1$,
which in turn translates to an upper bound on $v_t$
\begin{align}
|v_t| < 3.43\ \mbox{GeV},
\nonumber
\end{align}
and provides the main constraint on
the parameters $\lambda$ and $M_T$.
In Fig.~\ref{fig:TPlot1},
we plot the allowed regions on $M_T-\lambda$ plane
with $B_T=m_T^2=A^2_{\lambda}=(200\ \mbox{GeV})^2$, for various
values of $\tan\beta$ and $\mu$.
For small values of $\tan\beta$ and $\mu$,
$\alpha \Delta T$ is only viable with either small $\lambda$ or
a cancellation in the numerator of Eq.~\ref{eq:ObliqueTFormula}.
In Fig.~\ref{fig:TPlot2}, we plot the allowed region in $\mu-\tan\beta$ plane
for $\lambda=0.9$, $B_T=m_T^2=A^2_{\lambda}=(200\ \mbox{GeV})^2$,
for various values of $M_T$.
As expected, for larger values of $M_T$, there is a thicker band on
the $\mu-\tan\beta$ plane that is allowed.

We will quantify the degree of fine-tuning in the cancellation for
allowed $\alpha \Delta T$ in Sec.~\ref{sec:FineTuning}.
For now, we may estimate the fine-tuning along the
ideas of Athron et al.~\cite{Athron:2007ry}.
For example, with the parameters that require a fine-tuning
in $M_T$
\begin{align}
\tan\beta=3,\quad\lambda =0.9,\quad \mu =150\ \mbox{GeV},\nonumber\\
m_T^2=B_T=A_{\lambda}^2= (200\ \mbox{GeV})^2,
\label{eq:Texample}
\end{align}
we have viable $\Delta T$ in the regions
\begin{align}
250\ \mbox{GeV} < M_T < 375\ \mbox{GeV},\
\mbox{or}\ M_T> 3.0\ \mbox{TeV}.
\label{eq:Texmaple-range}
\end{align}
For $M_T$ below 3.0 TeV, we would typically
have unacceptably large $\Delta T$ that violates
precision electroweak constraints except in
the a small region of $M_T$ between
250 GeV and 375 GeV because of cancellations
in the numerator of Eq.~\ref{eq:ObliqueTFormula}.
If we sample $M_T$ at random in the range
between 0 and 3.0 TeV,
the only region
with viable $\Delta T$ is only 125(=375-250) GeV wide,
and we can thus estimate the fine-tuning as
\begin{align}
\frac{3.0\ \mbox{TeV}}{375\ \mbox{GeV}-250\ \mbox{GeV}}= 24,
\end{align}
so a cancellation of 1 part in 24 is required to have small $\alpha \Delta T$
for the parameters in Eq.~\ref{eq:Texample}.


\subsection{Neutralino and Charginos}
After EWSB, the neutralino ($\widetilde{N}$) and chargino ($\widetilde{C}$) mass matrices
are now extended with the triplet sector.
The mass matrix for the neutralinos in
the basis ($\widetilde{b}$, $\widetilde{w}^0$, $\widetilde{H}^0_d$
$\widetilde{H}^0_u$, $\widetilde{T}^0$) is given by
\newcommand{\fr}[2]{\frac{#1}{#2}}
\begin{align}
\mathcal{M}_{\widetilde{N}}=
\begin{pmatrix}
M_1 & 0 & -\fr{1}{2}g_1 v_d & \fr{1}{2}g_1 v_u & 0 \\
0 & M_2 & \fr{1}{2}g_2 v_d & -\fr{1}{2}g_2 v_u & 0 \\
-\fr{1}{2} g_1 v_d &  \fr{1}{2} g_2 v_d & 0 & -\mu_{\smbox{eff}} & \frac{1}{\sqrt{2}}\lambda v_u \\
\fr{1}{2} g_1 v_u &  -\fr{1}{2} g_2 v_u & -\mu_{\smbox{eff}} & 0 & \frac{1}{\sqrt{2}}\lambda v_d \\
0 & 0 & \frac{1}{\sqrt{2}}\lambda v_u & \frac{1}{\sqrt{2}}\lambda v_d & M_T
\end{pmatrix},
\label{eq:NMatrix}
\end{align}
where $M_1$ and $M_2$ are respectively the SUSY-breaking
bino and wino masses, and $\mu_{\smbox{eff}}$ is as defined
in Eq.~\ref{eq:mueff}.

For the charginos, in the basis
$\psi^{\pm}=(\widetilde{w}^+,\widetilde{H}_u^+,\widetilde{T}^+,
\widetilde{w}^-,\widetilde{H}_d^-,\widetilde{T}^-)$,
the chargino mass matrix  appears in the Lagrangian as
\begin{align}
\mathcal{L}\supset-\frac{1}{2}
(\psi^{\pm})^T
\begin{pmatrix} 0 & \mathcal{M}_{\widetilde{C}}^T \\
\mathcal{M}_{\widetilde{C}} & 0
\end{pmatrix}\psi^{\pm},
\end{align}
where
\begin{align}
\mathcal{M}_{\widetilde{C}}=
\begin{pmatrix}
M_2 & \fr{\sqrt{2}}{2}g_2 v_d & g_2 v_t&  \\
\fr{\sqrt{2}}{2} g_2 v_u& \mu_{\smbox{eff}}+\sqrt{2}\lambda v_t & -\lambda v_d&  \\
-g_2v_t & \lambda v_u & M_T
\end{pmatrix}.
\label{eq:CMatrix}
\end{align}

\section{Lightest $CP$-even Higgs Boson in the TESSM}
\label{sec:HiggsMass}
\subsection{Tree-Level Mass}
The lightest $CP$-even Higgs boson in TESSM
is a linear combination of the $CP$-even components of
the Higgs doublets $H_{u,d}$ and the neutral component
of the triplet $T^0$.
After EWSB, the squared-mass matrix for the
neutral scalar has the entries
\begin{align}
\left(\mathcal{M}^2_{a}\right)_{11}&=
\cosb^2 M_A^2+\sinb^2 M_Z^2 -\frac{\lambda}{\sqrt{2}}\tanb^{-1} v_t (M_T+A_{\lambda}),
\nonumber\\
\left(\mathcal{M}^2_{a}\right)_{12}&=
-\sinb\cosb(M_A^2+M_Z^2)+\cosb\sinb\lambda^2 v^2
+\frac{\lambda}{\sqrt{2}}v_t (M_T+A_{\lambda}),
\nonumber\\
\left(\mathcal{M}^2_{a}\right)_{13}&=
\lambda v\left(
\frac{1}{\sqrt{2}}(A_{\lambda}+M_T)\cosb+(\lambda v_t-\sqrt{2}\mu)\sinb
\right)
\nonumber\\
\left(\mathcal{M}^2_{a}\right)_{22}&=
\sinb^2 M_A^2+\cosb^2 M_Z^2 -\frac{\lambda}{\sqrt{2}}\tanb v_t (M_T+A_{\lambda}),
\nonumber\\
\left(\mathcal{M}^2_{a}\right)_{23}&=
\lambda v\left(
\frac{1}{\sqrt{2}}(A_{\lambda}+M_T)\sinb+(\lambda v_t-\sqrt{2}\mu)\cosb
\right)
\nonumber\\
\left(\mathcal{M}^2_{a}\right)_{33}&=
M_T^2+m_T^2+B_T+\frac{1}{2}\lambda^2 v^2,
\end{align}
where $v_t$ should be considered as a function of the
input parameters via the minimization condition
in Eq.~\ref{eq:vtCond} and we define $M_A$ as in the case of
the MSSM
\begin{align}
M^2_A\equiv 2 \frac{B_\mu}{\sin 2\beta}.
\end{align}

\begin{figure}[h!t]
\begin{center}
\includegraphics[width=4.25in]{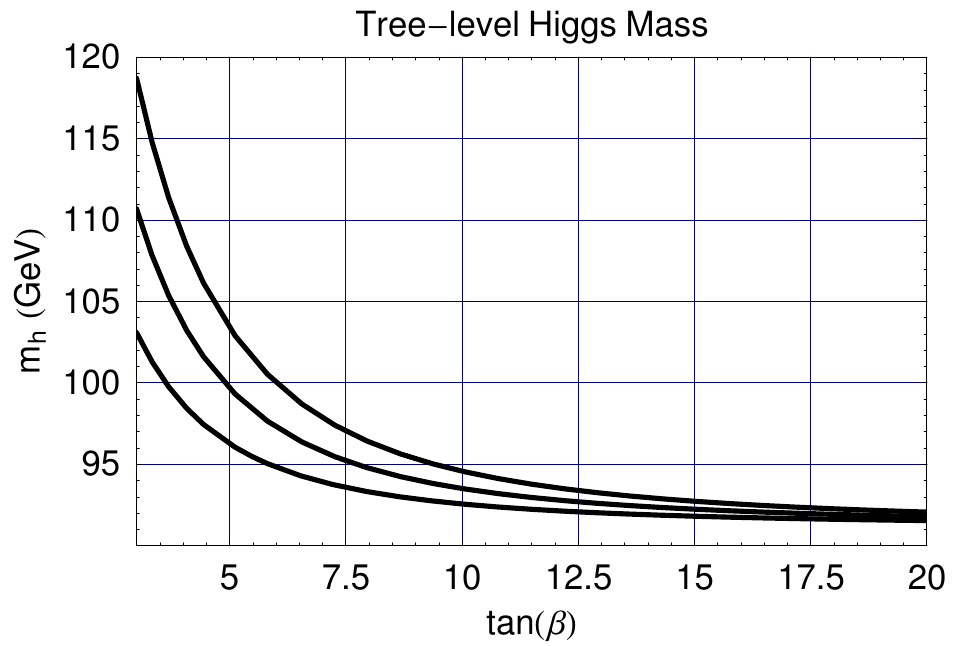}
\caption{Tree-level upper bound on the mass of the lightest
$CP$-even Higgs boson as given by Eq.~\ref{eq:TreeMassBound} as a
function of $\tan\beta$ for $\lambda=0.7$ (lowest line), 0.8
(middle line), and 0.9 (top line) } \label{fig:TreePlot}
\end{center}
\end{figure}

As in the case of the NMSSM, the lightest mass-squared eigenvalue
is bounded by the lightest eigenvalue of top-left 2$\times$2
block of the mass matrix,
\begin{align}
m^2_h \leq M^2_Z
\left(\cos 2\beta+\frac{2\lambda^2}{g_2^2+g_1^2}\sin 2\beta
\right).
\label{eq:TreeMassBound}
\end{align}
In Fig.~\ref{fig:TreePlot}, we plot this tree-level upper bound
as a function of $\tan\beta$ for $\lambda=0.7$, 0.8, and 0.9.
For $\lambda=0.9$, it is possible to obtain a tree-level Higgs mass larger than
100 GeV for $\tan\beta\lesssim 6$, and even satisfy the LEP2 bounds at
tree-level for small $\tan\beta\lesssim 3.5$.

\begin{figure}[h!t]
\begin{center}
\includegraphics[width=\picwidth]{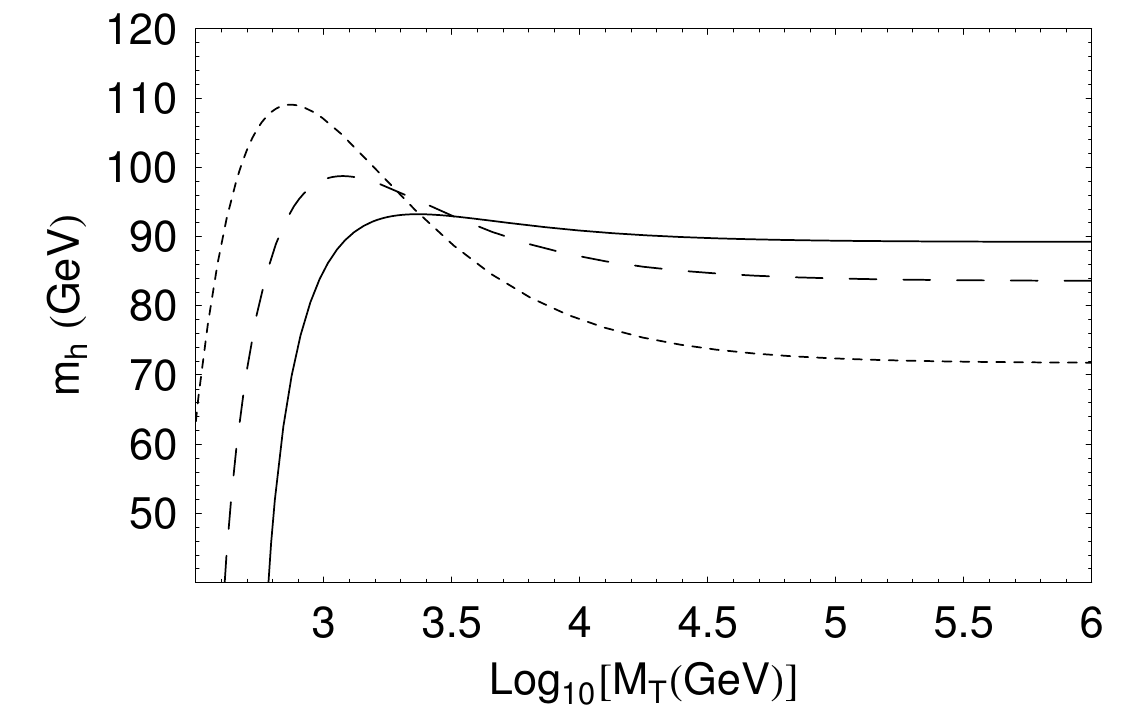}
\includegraphics[width=\picwidth]{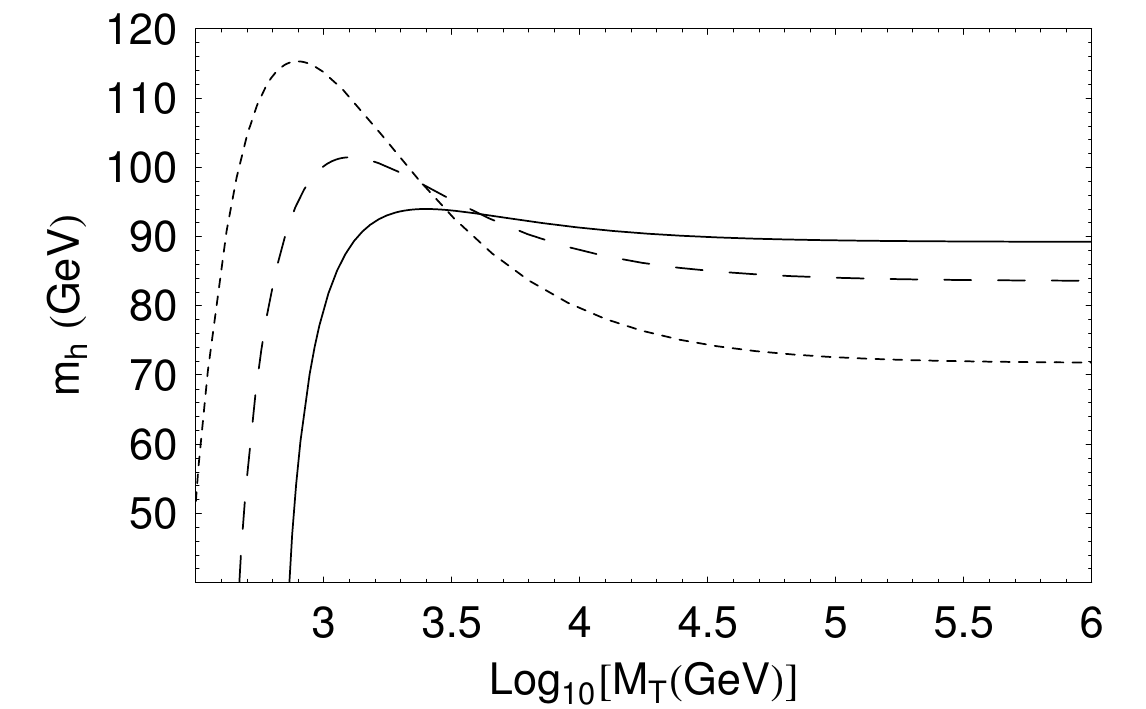}
\caption{Tree-level mass of the lightest $CP$-even Higgs boson as
a function $M_T$ for $\lambda=0.8$ (left) and $\lambda=0.9$ (right).
The other parameters are kept fixed as
$A_{\lambda}=m_T^2=B_T=0$, $\mu=200$ GeV, and $M_A=300$ GeV.
The
three curves have values of $\tan\beta$ of 10 (solid), 5 (dashed),
and 3 (dotted). For each case, we see decoupling in large $M_T$,
and the limiting value agrees with the MSSM result. }
\label{fig:DePlot}
\end{center}
\end{figure}

We can also see the MSSM limit in the mass matrix when the triplet decouples with
fixed $\lambda$.
In the limit $M_T\rightarrow\infty$, keeping all other parameters fixed,
we have
\begin{align}
M_T v_t \rightarrow -\frac{\lambda v^2}{2\sqrt{2}}\sin 2\beta,
\end{align}
and the mass matrix has the form
\begin{align}
\mathcal{M}^2_a\stackrel{M_T\rightarrow\infty}{\longrightarrow}
\begin{pmatrix} \mathcal{M}^2_{\smbox{MSSM}}+\Delta\mathcal{M}^2
& \epsilon \\ \epsilon^{T} & M_T^2
\end{pmatrix},
\label{eq:MA-approx}
\end{align}
where $\mathcal{M}^2_{\smbox{MSSM}}$ is the 2$\times$2 MSSM mass matrix for the $CP$-even
Higgs bosons, and
\begin{align}
\Delta\mathcal{M}^2&=\lambda^2\frac{v^2}{2}
\begin{pmatrix}
\cosb^2 & \cosb\sinb \\ \cosb\sinb & \sinb^2
\end{pmatrix},\\
\epsilon &= \frac{\lambda}{\sqrt{2}} M_T v
\begin{pmatrix}
\cosb \\ \sinb
\end{pmatrix}.
\end{align}
Integrating out the third row and column of the mass matrix in
Eq.~\ref{eq:MA-approx}, the effective top-left 2$\times$2 sub-matrix
becomes
\begin{align}
\mathcal{M}^2_{\smbox{eff}}&=
\mathcal{M}^2_{\smbox{MSSM}}+\Delta\mathcal{M}^2
-\epsilon (M_T^2)^{-1}\epsilon^{T}
+\mathcal{O}\left(\frac{\epsilon^3}{M_T^4}\right)\nonumber\\
&=\mathcal{M}^2_{\smbox{MSSM}}+\mathcal{O}\left(\frac{\lambda^3 v^3}{M_T}\right),
\end{align}
and we recover the MSSM limit as $M_T\rightarrow\infty$.
In Fig.~\ref{fig:DePlot}, we show this decoupling behavior by plotting
the tree-level mass of the lightest Higgs boson as a function of $M_T$ for various
values of $\tan\beta$, and see that, for $M_T \gtrsim 10^4$ GeV,
we recover the MSSM results.

\subsubsection{Numerical Results}
In this subsection, we numerically evaluate the mass of the lightest
$CP$-even Higgs boson at tree-level.
With the minimization conditions, we can take as input parameters
\begin{align}
\tan\beta,\, \mu,\, M_A,\,
\lambda,\, M_T,\, m_T^2,\, B_T,
\end{align}
and fix $m_{H_u}^2$, $m_{H_d}^2$ and $v_t$ by solving the
minimization conditions with the experimental inputs of
$M_Z=91.19$ GeV and the gauge couplings $g_2(M_Z)\simeq 0.65$,
$g_1(M_Z)\simeq 0.36$ (this fixes $v\simeq 245$ GeV).
We discard sets of input parameters that give large $v_t$
inconsistent with electroweak constraints.
For all our numerical studies, we analyze two the cases of
$\lambda=0.8$ and $\lambda=0.9$, and
scan the parameter space in the range
\begin{eqnarray}
3\leq&\tan\beta &\leq 30,\nonumber\\
100\ \mbox{GeV}  \leq&\mu, M_A &\leq 500\ \mbox{GeV},\nonumber\\
300\ \mbox{GeV}  \leq& M_T  &\leq 1000\ \mbox{GeV},\nonumber\\
-2000\ \mbox{GeV} \leq& A_{\lambda}  &\leq 2000\ \mbox{GeV},\nonumber\\
-(1000\ \mbox{GeV})^2\leq& m_T^2,B_T&\leq (1000\ \mbox{GeV})^2.
\label{eq:ScannedRange}
\end{eqnarray}
The range of $\tan\beta$ is chosen so that the bottom Yukawa coupling
is smaller than the top Yukawa coupling, and we can neglect the bottom Yukawa coupling
when we study the production and decay properties of the lightest $CP$-even
Higgs boson.

Since solutions to the minimization conditions only guarantee an extremum,
we only keep solutions that give a local minimum by checking that
all scalar masses are positive at the desired vev.
We also discard any points
that give unacceptably large $\alpha\Delta T$ or contain charged scalar particles
lighter than 100 GeV.

\begin{figure}[h!t]
\begin{center}
\includegraphics[width=\picwidth]{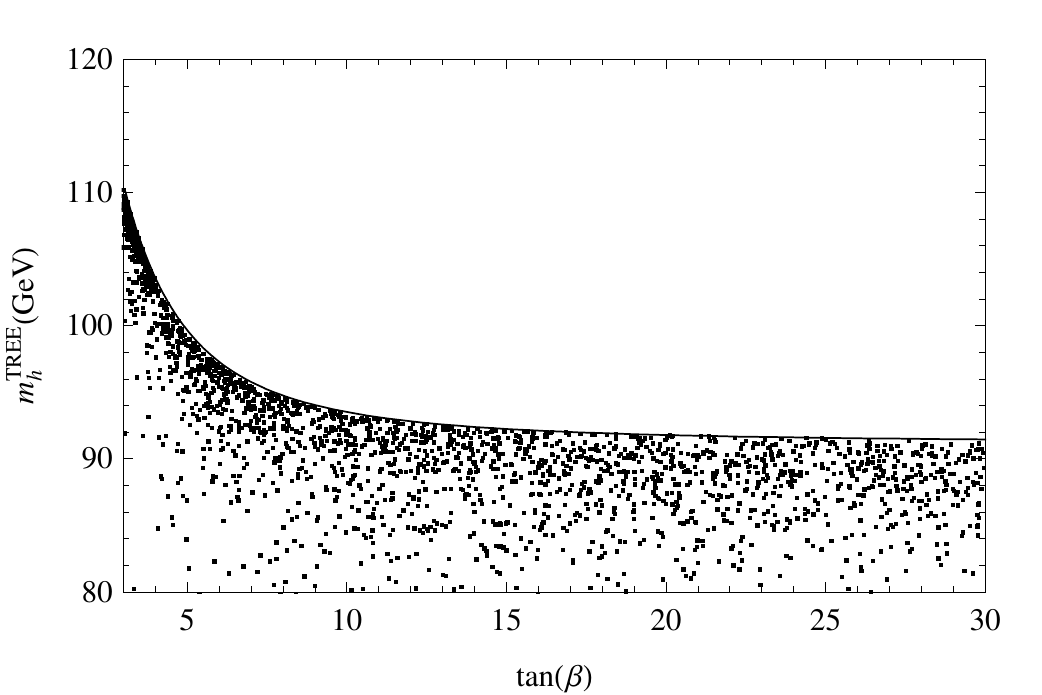}
\includegraphics[width=\picwidth]{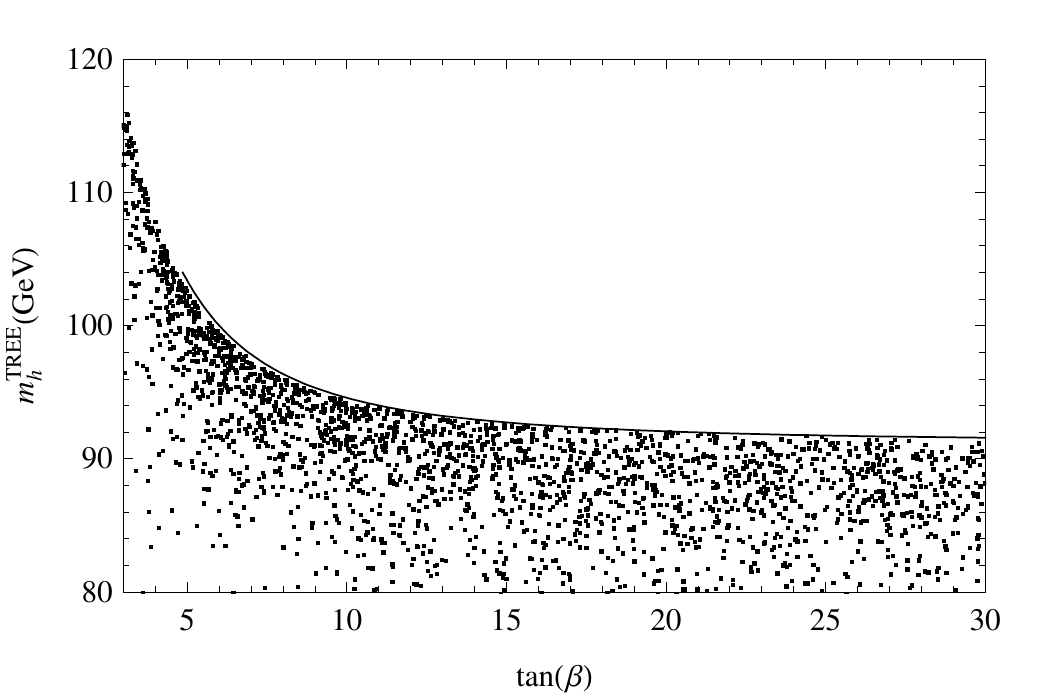}
\caption{
Tree-level mass of the lightest $CP$-even Higgs boson as a function
$\tan\beta$ scanned over the parameter space as listed in
Eq.~\ref{eq:ScannedRange}.
The plot on the left (right) has $\lambda=0.8$ ($\lambda=0.9$), and the
top curve indicates the upper bound as computed from Eq.~\ref{eq:TreeMassBound}.
}
\label{fig:NTreePlot}
\end{center}
\end{figure}

For the range of parameters listed in Eq.~\ref{eq:ScannedRange},
we show the mass of the lightest $CP$-even boson as a function
of $\tan\beta$ in Fig.~\ref{fig:NTreePlot}.
In Fig.~\ref{fig:NTreePlot}, we also plot the upper bound of the tree-level Higgs mass
given in Eq.~\ref{eq:TreeMassBound}.
The plots show that we can indeed achieve large (greater than $M_Z$)
tree-level Higgs mass
with large $\lambda$, and we can even satisfy LEP2 bounds at tree-level
for small values of $\tan\beta$ ($\tan\beta\lesssim 3.5$) when $\lambda=0.9$.

\subsection{Mass at the One-Loop Level}
Since the lightest $CP$-even Higgs boson is a linear
combination of $a^{\prime}_i$ for $i=u,d,t$, we will
construct the Coleman-Weinberg (CW) potential \cite{Coleman:1973jx} only for
the fields $a_i$, and extract the corrections to $m_h^2$
from the CW potential.
Furthermore, we will make the two following assumptions:
\begin{itemize}
\item
We assume that both the s-top masses are close to the top-quark
mass, and the famous $\mathcal{O}(y_t^4)$ contributions in the
MSSM are small.
In other words, we only consider the corrections from
the Higgs boson, neutralino, and chargino sectors.
These contributions are dominated by the coupling $\lambda$ and
the SUSY-breaking parameters in the triplet sector.
Our results will show that these contributions
are sufficient to satisfy the LEP2 bounds on the Higgs
mass, and we do not need large contributions from
the top--s-top sector
as in the case of the MSSM.
\item
In the neutralino and chargino mass matrices, we ignore
the mixing induced by gauge interactions.
This removes dependencies on the bino and wino SUSY-breaking masses
in our analysis as
we do not include the bino and wino states,
and we expect
their contributions to be small
when $M_{1,2}\sim M_Z$.
(If we include the bino and wino states, we would also have to
include the corresponding superpartners in the $W$ and $Z$ gauge
bosons.)
\end{itemize}

The Coleman-Weinberg potential is given by
\begin{align}
V_{\smbox{CW}}=\frac{1}{64\pi^2}
\mbox{STr}\left[ \mathcal{M}^4
\left(\ln\frac{\mathcal{M}^2}{\mu_r^2}-\frac{3}{2}\right)\right],
\end{align}
where $\mathcal{M}^2$ are field-dependent mass matrices
in which the fields are not replaced with their vev's,
$\mu_r$ is the renormalization scale,
and the supertrace includes a factor of $(-1)^{2J}(2J+1)$
so that fermions contribute oppositely to bosons, and
the spin degrees of freedoms are appropriately summed over.
Since here we are only interested in the CW potential
of the fields $a_i$ that always appear in the combination
$(a^{\prime}_i + v_i)$, the field-dependent matrices for the
charginos and neutralinos are simply those in
Eqs.~\ref{eq:NMatrix} and \ref{eq:CMatrix} with
the vevs $v_i$ replaced by the corresponding fields
$a_i$.

For the scalars, the naive replace-vev-by-field method fails and
we need to distinguish between
the contributions from the minimization conditions
and those from the replacement of the fields with their
corresponding vevs.
For example, while the (11)-element of the mass matrix of
the $CP$-even
neutral Higgs boson is
\begin{align}
\left(\mathcal{M}^2_{a}\right)_{11}&=
\cosb^2 M_A^2+\sinb^2 M_Z^2 -\frac{\lambda}{\sqrt{2}}\tanb^{-1} v_t (M_T+A_{\lambda}),\\
&=B_{\mu}\frac{v_d}{v_u}+\frac{1}{4}(g^2_2+g_1^2)v_u^2-\frac{\lambda}{\sqrt{2}}v_t
(M_T+A_{\lambda})\frac{v_d}{v_u},
\label{eq:CWex01}
\end{align}
it is incorrect to have the field-dependence
\begin{align}
\left(\mathcal{M}^2_{a}\right)_{11}&\neq
B_{\mu}\frac{a_d}{a_u}+\frac{1}{4}(g^2_2+g_1^2)a_u^2-\frac{\lambda}{\sqrt{2}}a_t
(M_T+A_{\lambda})\frac{a_d}{a_u},
\end{align}
because some of the vev-dependence in Eq.~\ref{eq:CWex01} comes from
the minimization conditions Eq.~\ref{eq:Min-MHU}.
The correct
field-dependent (11)-element of the $CP$-even
neutral Higgs boson is
\begin{align}
(\mathcal{M}^2_{a})_{11}
&=
m^2_{H_u}+\mu^2+\frac{g_2^2+g_1^2}{8}(3a_u^2-a_d^2)
+\frac{\lambda^2}{2}(a_d^2+a_t^2)-\sqrt{2}\lambda \mu a_t,
\end{align}
where $m^2_{H_u}$ is related to the vev's (but not the fields)
through the minimization condition in Eq.~\ref{eq:Min-MHU}.
In Appendix \ref{app:Field-dep-masses},
we give the field-dependent mass matrices used in
the calculation of the Coleman-Weinberg potential.

Since the analytical results for the mass eigenvalues
of the field-dependent matrices are complicated,
we will compute the Higgs mass
numerically.
The one-loop mass matrix can be
extracted from the Coleman-Weinberg potential
by numerically evaluating the derivatives of the mass eigenvalues
with respect to the fields
about the vevs \cite{Elliott:1993bs} (dropping
the pre-factor from the supertrace for convenience)
\begin{align}
(\Delta\mathcal{M}^2_{a})_{ij}
&=\left.\frac{\partial^2 V_{\smbox{CW}}(a)}{\partial a_i\partial a_j}\right|_{\smbox{vev}}
-\frac{\delta_{ij}}{\vev{a_i}}\left.\frac{\partial V_{\smbox{CW}}(a)}{\partial a_i}\right|_{\smbox{vev}}
\label{eq:CWex02}\\
&=\sum\limits_{k}\frac{1}{32\pi^2}
\frac{\partial m^2_k}{\partial a_i}
\frac{\partial m^2_k}{\partial a_j}
\left.\ln\frac{m_k^2}{\mu_r^2}\right|_{\smbox{vev}}
+\sum\limits_{k}\frac{1}{32\pi^2}
m^2_k\frac{\partial^2 m^2_k}{\partial a_i\partial a_j}
\left.\left(\ln\frac{m_k^2}{\mu_r^2}-1\right)\right|_{\smbox{vev}}
\nonumber\\
&\quad-\sum\limits_{k}\frac{1}{32\pi^2}m^2_k
\frac{\delta_{ij}}{\vev{a_i}}
\frac{\partial m^2_k}{\partial a_i}
\left.\left(\ln\frac{m_k^2}{\mu_r^2}-1\right)\right|_{\smbox{vev}}
\label{eq:CWex03}
\end{align}
where the second term in Eq.~\ref{eq:CWex02} takes into account the shift in
the minimization conditions, and $\{m^2_k\}$ is the set of
mass eigenvalues that enter the Coleman-Weinberg potential.
Our set of $\{m^2_k\}$
includes the eigenvalues of the mass matrices of
the $CP$-even, $CP$-odd, and charged Higgs bosons, as well as the
neutralinos and charginos mass matrices.
These field-dependent mass matrices are given
in Appendix \ref{app:Field-dep-masses}.

\subsubsection{Numerical Results}
\begin{figure}[h!t]
\begin{center}
\includegraphics[width=\picwidth]{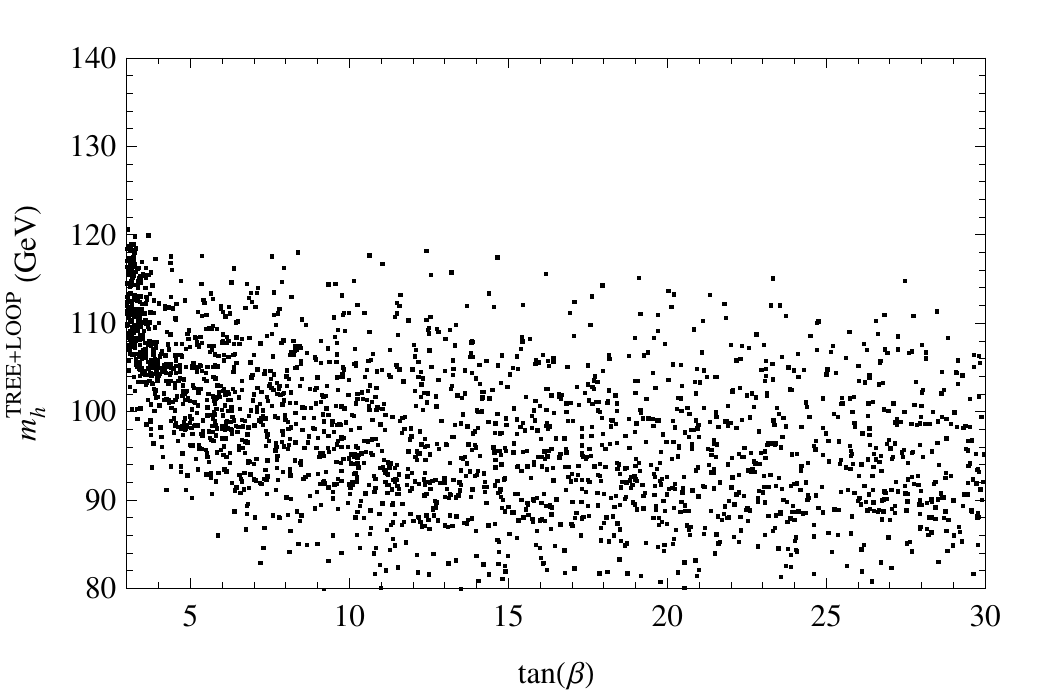}
\includegraphics[width=\picwidth]{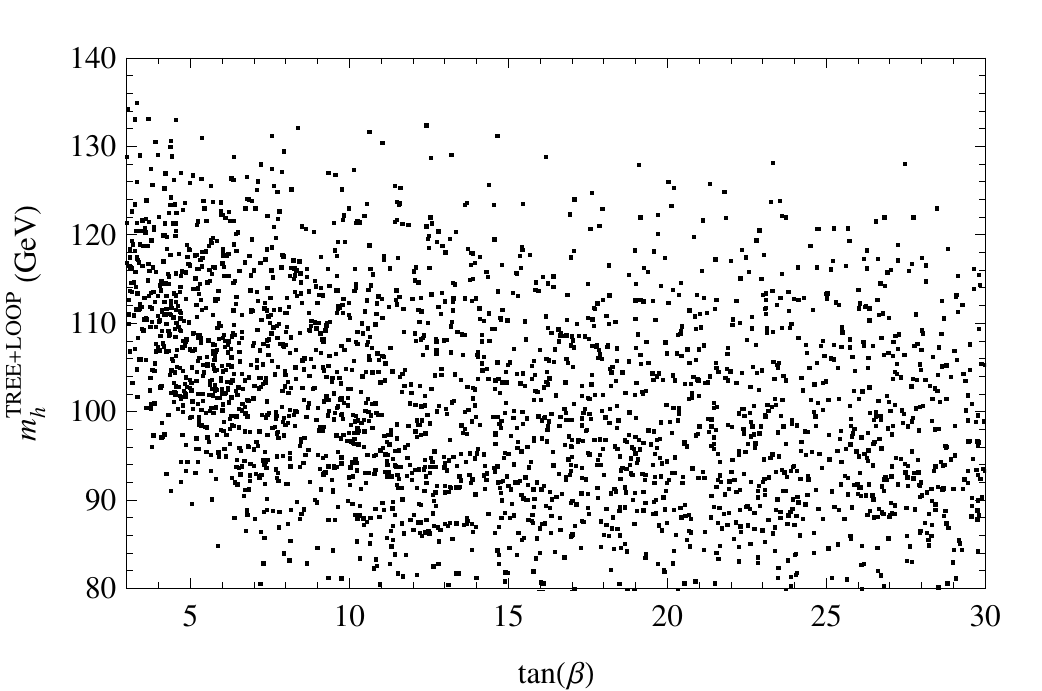}
\caption{
Mass of the lightest $CP$-even Higgs boson,
including one-loop contributions from
the triplet sector, as a function
$\tan\beta$, scanned over the parameter space as listed in
Eq.~\ref{eq:ScannedRange}.
The plot on the left has $\lambda=0.8$, and
the plot on the right has $\lambda=0.9$.
The input parameters of the individual points are
the same as those that give rise to the points
shown in the corresponding plot of Fig.~\ref{fig:NTreePlot}.
}
\label{fig:NLoopPlot}
\end{center}
\end{figure}
\begin{figure}[h!t]
\begin{center}
\includegraphics[width=\picwidth]{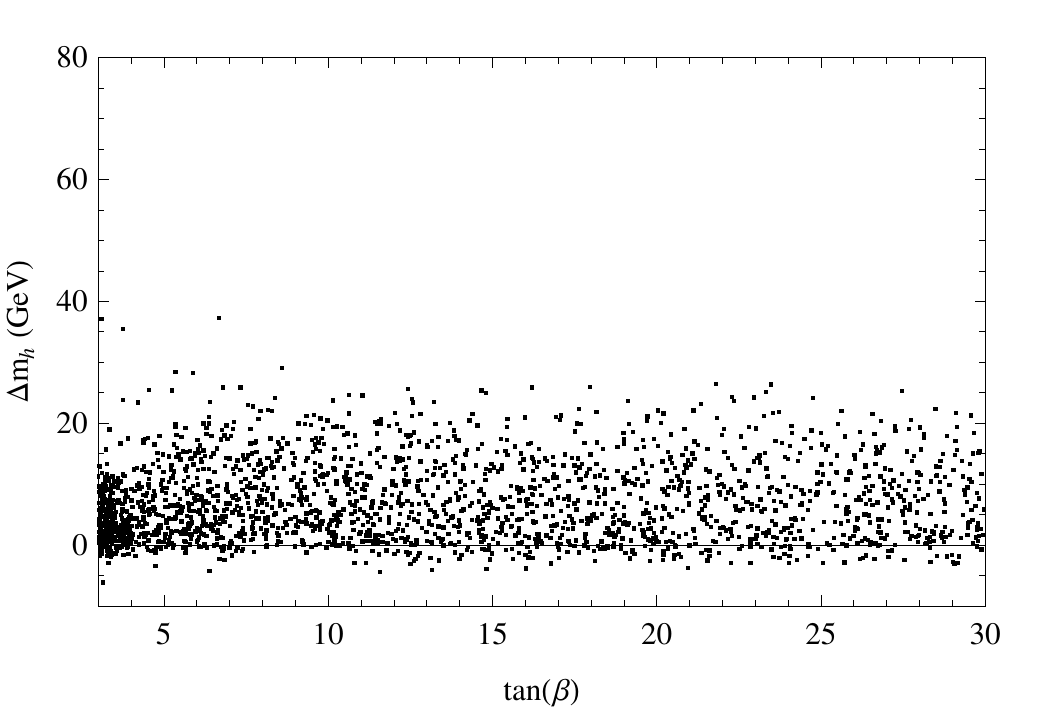}
\includegraphics[width=\picwidth]{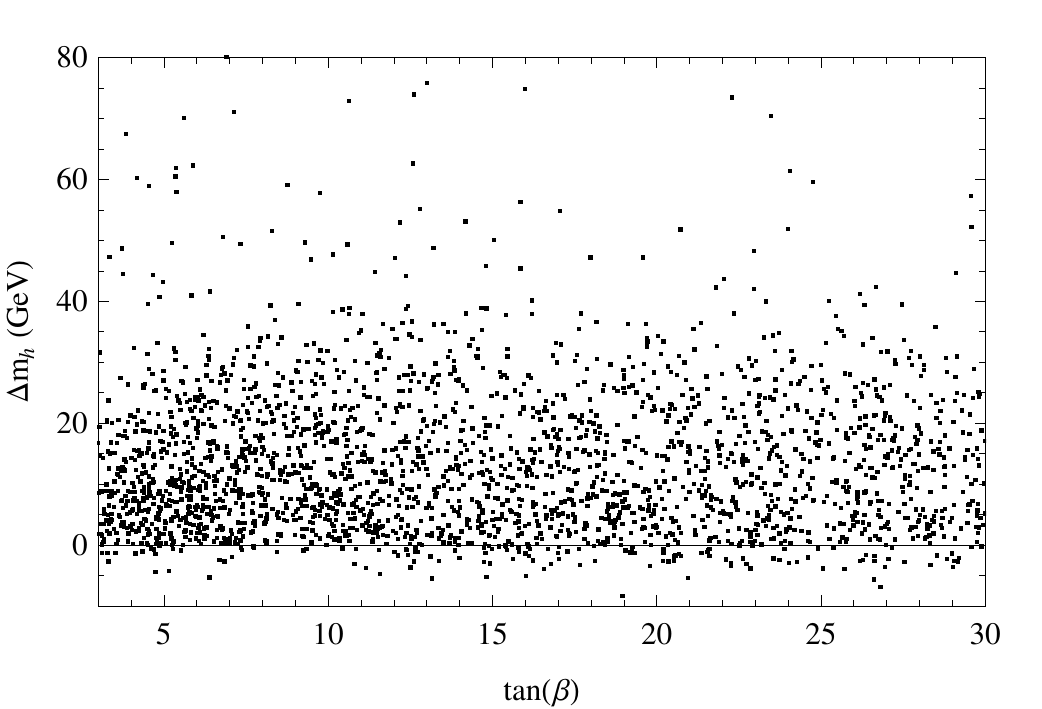}
\caption{
Difference between the mass of the lightest
$CP$-even Higgs boson with and without
the one-loop contribution from the triplet sector
for the points shown in the
corresponding plots of Figs.~\ref{fig:NLoopPlot}
and \ref{fig:NTreePlot}.
The plot on the left has $\lambda=0.8$, and
the plot on the right has $\lambda=0.9$.
}
\label{fig:NDiffPlot}
\end{center}
\end{figure}

We numerically compute the mass of the lightest $CP$-even Higgs boson
to one-loop using the Coleman-Weinberg potential
for the parameter space in Eq.~\ref{eq:ScannedRange}.
For the same input parameters
that give rise to the tree-level results shown in Fig.~\ref{fig:NTreePlot},
we show the corresponding Higgs mass computed to one loop
in Fig.~\ref{fig:NLoopPlot}, and the difference between
the loop-level and tree-level masses in Fig.~\ref{fig:NDiffPlot}.
We use the value of $M_T$ as the renormalization scale $\mu_r$ that
enters the Coleman-Weinberg potential.
From these plots, we see that the triplet sector can give a large contribution to
the mass of the lightest $CP$-even Higgs boson, and we can
satisfy the LEP bounds without the s-top contributions for all values
of $\tan\beta$ in our scanned range.

\begin{table}[h]
\begin{center}
\caption{Sample Higgs spectra.
All dimensionful parameters
are in units of GeV, except for $m_T^2$ and $B_T$,
which are in units of (GeV)$^2$.
The definitions of fine-tuning $f_T$,
$\kappa_T$, and $\kappa_T^{\prime}$
are given in, respectively,
Eqs.~\ref{eq:HiggsFineTune},
\ref{eq:TripletFineTune}, and
\ref{eq:TripletFineTune2}.
The value of $f_T$ indicates
the percent change in $v^2$
induced by a 1\% change in $m^2_{H_u}$
at a fundamental scale of SUSY-breaking,
and
the value of $\kappa_T$ indicates
the percent change in $\Delta T$
induced by a 1\% change in $M_T$.
The measure $\kappa^{\prime}_T$ is
only applicable to Points 1 and 2, and shows that
there is a cancellation of one part in 23 to
give a viable value of $\Delta T$.
Points 1 and 2 show examples of input parameters
that give a viable Higgs mass with a small fine-tuning
in the electroweak sector.
Points 3 and 4 differ only in $\lambda$ and are samples that give large
Higgs masses of about 120 GeV (for $\lambda = 0.8$)
and about 135 GeV (for $\lambda = 0.9$).
Points 5 and 6 have large $\tan\beta(\geq 20)$ and
$m_h\sim M_Z$ at tree-level, but there
are large radiative corrections to have viable Higgs
masses at one-loop.
}
\label{tb:SamplePoints}
\vspace{0.125in}
\begin{tabular}{|c|c|c||c|c||c|c|}
\hline & Point 1 & Point 2 & Point 3 & Point 4 & Point 5 & Point 6\\
\hline $\tan\beta$ & 3.20 & 3.20 & 3.20 & 3.20 & 20.0 & 27.7\\
\hline $\mu$ & 270 & 270 & 400 & 400 & 200 & 165 \\
\hline $M_A$ & 430 & 430 & 280 & 280 & 300 & 410\\
\hline $\lambda$ & 0.8 & 0.9 & 0.8 & 0.9 & 0.8 & 0.9\\
\hline $M_T$ & 370 & 400 & 400 & 400 & 350 & 330 \\
\hline $m_T^2$ & (500)$^2$ & (280)$^2$ & (1970)$^2$ & (1970)$^2$ & (1600)$^2$ & $(1500)^2$\\
\hline $A_{\lambda}$ & 600 & 460 & 1860 & 1860 & 1800 & 1300 \\
\hline $B_T$ & (400)$^2$ & (400)$^2$ & (1730)$^2$ & (1730)$^2$ & (500)$^2$ & $(1400)^2$\\
\hline
\hline $m_h^{\smbox{Tree}}$ & 108 & 113 & 105 & 111 & 88 & 90\\
\hline $m^{\smbox{Tree+Loop}}_h$ & 114 & 117 & 122 & 137 & 114 & 121 \\
\hline
\hline $f_T$ & 33 & 19 & 399 & 505 & 315 & 271\\
\hline $\kappa_T$ & 33 & 11 & 0.8 & 0.8 & 0.5 & 0.3\\
\hline $\kappa^{\prime}_T$ & 4.7 & 6.4 &  &  &  & \\
\hline
\end{tabular}
\end{center}
\end{table}

In Table~\ref{tb:SamplePoints}, we give some sample points in our scan.
Points 1 and 2 are sample points that have small fine-tuning (as will be defined later
in Eq.~\ref{eq:HiggsFineTune}).
The TESSM can achieve small fine-tuning because the Higgs mass can be large
at tree-level and does not require large contributions from
radiative corrections.
Points 3 and 4 are samples of the points with the largest Higgs
masses (and therefore fine-tuning) in our scanned range of
parameter space, as evident in the values of SUSY-breaking
parameters $m_T^2$, $A_{\lambda}$, and $B_T$ being near the
boundary of the scanned range.
Points 5 and 6 are samples of points having a large $\tan\beta(\geq 20)$,
where the tree-level Higgs mass is only slightly larger than $M_Z$,
and there is a significant one-loop contribution, and, correspondingly,
large fine-tuning.

\subsection{Collider Signatures of the Lightest $CP$-even Higgs Boson in TESSM}
With large $\lambda$ in both the TESSM and the NMSSM, we do not
require heavy s-tops.
In these cases, the gluon-gluon
fusion production of the lightest $CP$-even Higgs boson,
$\sigma(gg\rightarrow h)$, and its diphoton partial decay width,
$\Gamma(h\rightarrow\gamma\gamma)$, can be
very different from the MSSM because
these processes involve s-top loops.
In this section we perform a simplified analysis showing that
in the TESSM there may be a gluo-philic Higgs boson whose
gluon-gluon fusion production cross section can be larger than that of the SM
by a factor of 1.8.
As stated in the introduction, our discussions of the gluon-gluon fusion production
rely only on the existence of light s-tops and the
minimal color sector of the MSSM, and are therefore applicable to
any extensions of the MSSM that solves the little hierarchy problem
without invoking additional colored states.
For the diphoton partial decay width, there are several
sources of suppression, and we may have a partial decay width
that is about 0.8 times that in the SM.

Of course, at the LHC the relevant quantity is
the product of the gluon-gluon fusion production cross section
and the diphoton branching ratio
\begin{align}
\sigma(gg\rightarrow h)
\mbox{Br}(h\rightarrow\gamma\gamma),\nonumber
\end{align}
and a more complete analysis would have to take into account
the effects of light s-tops to all the decay channels
as well
as the large, higher-loop corrections from QCD and
large couplings.
We leave the complete analysis of
the Higgs production and decay for future work.

The well-known formula for the decay width of a real scalar
decaying into two photons can be found in Gunion et
al.~\cite{Gunion:1984yn}.
This formula is also presented in the Appendix \ref{app:di-photon-width}.

\subsubsection{Gluo-philic Higgs boson}
\begin{figure}[h!t]
\begin{center}
\includegraphics[width=4.25in]{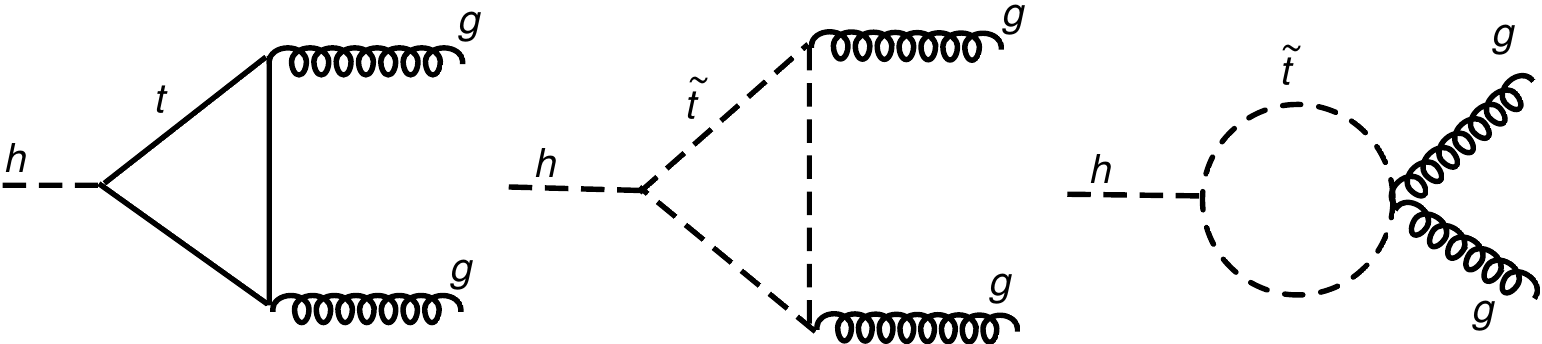}
\caption{
The diagrams that contribute to
the amplitude $\mathcal{A}(h\rightarrow gg)$
in the TESSM.
}
\label{fig:feyn-digluon}
\end{center}
\end{figure}

In the SM, ignoring all the Yukawa couplings except for the top
Yukawa coupling, the process $h\rightarrow gg$ proceeds only
through a top-quark loop.
In the MSSM, we have additional
contributions from the s-tops (see Fig.~\ref{fig:feyn-digluon}), as well
as all the other s-quarks through $D$-term
interactions of the form $h\widetilde{q}^{\ast}\widetilde{q}$
with coupling of the order $M_Z$ .
To simplify our analysis,
we will ignore the $D$-term interactions except
those in the s-top sector, but we note that these contributions
can be important
when there are light s-quarks
and must be taken into account in a full analysis.

In the MSSM with small s-top mixing,
the s-top contributions interfere constructively
with the top-quark contribution for the gluon-gluon fusion
production cross section \cite{Dawson:1996xz}\cite{Harlander:2004tp}.
However, with small s-top mixing, the s-tops need to be
heavy to satisfy the LEP bounds on the Higgs mass, and
the s-top contributions decouple.
(With large-stop mixing, it is possible to have
s-top contributions that interfere
destructively with the top-quark contribution,
leading to a gluo-phobic Higgs boson.)

\begin{figure}[h!t]
\begin{center}
\includegraphics[width=4.25in]{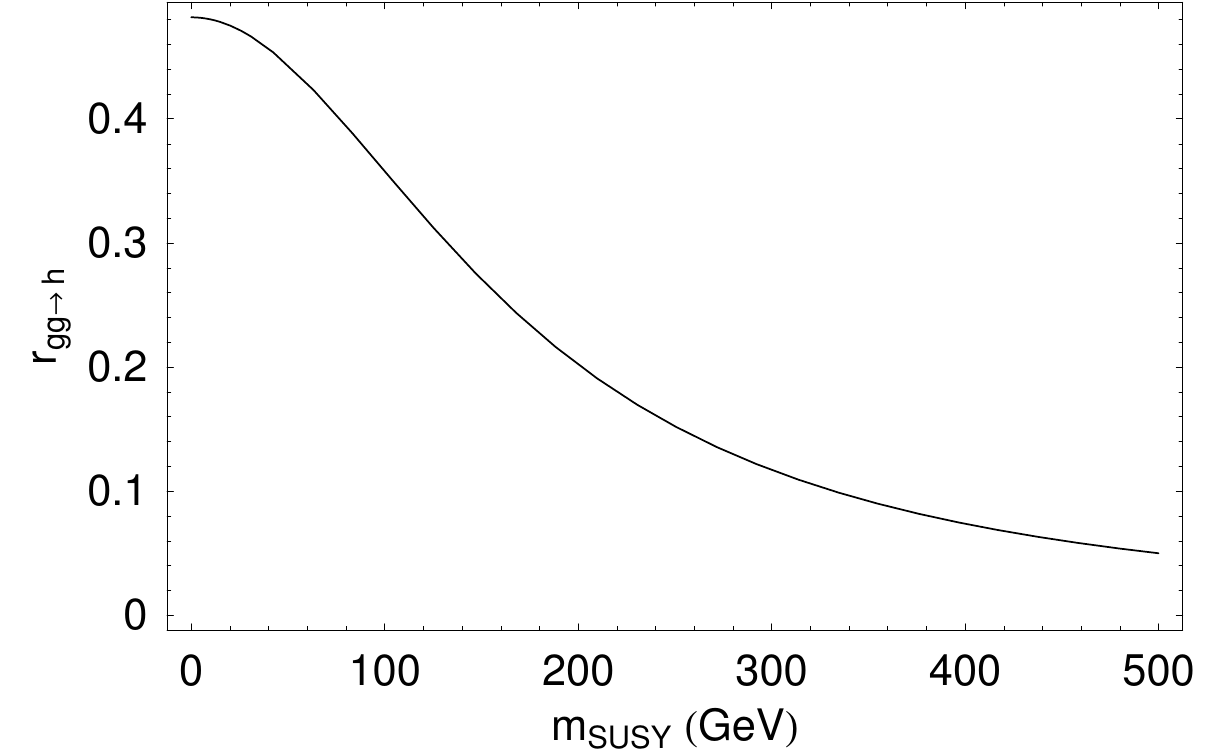}
\caption{ The amplitude $\mathcal{A}(gg\rightarrow h)$ through
s-top loops normalized with respect to the amplitude through
top-quark loop, as a function of a common s-top soft mass
$m_{\widetilde{t}}$, assuming no mixing in the s-top sector. }
\label{fig:r-h-to-gg}
\end{center}
\end{figure}

In the TESSM and NMSSM, we can have light s-tops at the expense of
perturbativity at the GUT scale, and
a large enhancement in the production rate.
Assuming large $\tan\beta$ so that $v\simeq v_u$ and there is no s-top mixing, and
approximating the lightest $CP$-even Higgs boson $h$ as being
dominantly composed of $a^{\prime}_u$
(the $CP$-even component of $H_u$), we have the interactions
\begin{align}
-\mathcal{L}\supset
\frac{y_t}{\sqrt{2}}\,h\, \overline{t}t
+
\left(y_t^2+\frac{1}{12}g_1^2-\frac{1}{4}g_2^2\right)\,v\,h \widetilde{Q}^{\ast}_3\widetilde{Q}_3
+
\left(y_t^2-\frac{1}{3}g_1^2\right)\,v\,h \widetilde{\overline{U}}^{\ast}_3\widetilde{\overline{U}}_3,
\end{align}
where $t$ is the top-quark, and $\widetilde{Q}_3$
($\widetilde{\overline{U}}_3$) is the superpartner to the
left-(right-)handed component of the top-quark.
From Eqs.~\ref{eq:Di-photon-formula},
\ref{eq:diphoton-to-digluon}, and \ref{eq:sigma-gamma-conv}, the
ratio of the amplitudes $\mathcal{A}(gg\rightarrow h)$ due to the
s-top s-quarks and top-quark is then
\begin{align}
r_{gg\rightarrow h}\equiv\frac{\mathcal{A}_{\widetilde{t}}}{\mathcal{A}_t}
=\frac{m_t^2+\tfrac{1}{4}\left(\tfrac{1}{6}g_1^2-\tfrac{1}{2}g_2^2\right)v^2}{m^2_{\widetilde{Q}_3}+m_t^2}
\frac{F_{\widetilde{Q}_3}}{F_t}
+\frac{m_t^2-\tfrac{1}{6}g_1^2v^2}{m^2_{\widetilde{\overline{U}}_3}+m_t^2}\frac{F_{\widetilde{\overline{U}}_3}}{F_t},
\end{align}
where $m^2_{\widetilde{Q}_3}$ and $m^2_{\widetilde{\overline{U}}_3}$ are SUSY-breaking
soft masses of the corresponding s-tops, and we have used the relationships
$m_t=(\sqrt{2})^{-1}y_t v_u$.
In Fig.~\ref{fig:r-h-to-gg}, we plot $r_{gg\rightarrow h}$ as a function
of a common soft s-top mass
$m^2_{\widetilde{Q}_3}=m^2_{\widetilde{\overline{U}}_3}=m^2_{\smbox{SUSY}}$,
assuming $m_h=114$ GeV.
Since the s-top mass eigenvalues in this simplified analysis are given by
\begin{align}
m^2_{\widetilde{t}} = m^2_{\smbox{SUSY}} + m_t^2,
\end{align}
and the current searches limit the
s-tops masses to be greater than 120 GeV \cite{Yao:2006px},
we can have $m_{\smbox{SUSY}}\sim 0$ (so $m_{\tilde{t}}=m_t$) and
$r_{gg\rightarrow h}$ can be as large as 0.48.
This gives a gluo-philic Higgs boson whose production cross section
via gluon-gluon fusion may be enhanced relative to the SM prediction
by a factor of $(1.48)^2\sim 2.2$.

Imposing perturbativity at the GUT scale, we can have a milder gluo-philic Higgs boson
when one of the s-top is light (the other
is required to be heavy to have a Higgs mass satisfying
the LEP2 bounds).
However, when only one s-top is light, the enhancement
in the gluon-gluon fusion production cross section is only
a factor of $(1+0.5\times 0.48)^2\sim 1.5$ larger than that of
the SM.

\subsubsection{Diphoton Decay of the Higgs boson}
In the SM, the diphoton decay of the Higgs boson proceeds through
$W$-boson loop in addition to top-quark loop, and the contribution
from the top-quark destructively interferes with the dominant
$W$-boson contribution.
In the MSSM, we have additional contributions from the s-tops and
charginos (the corresponding superpartners of the top-quark and
$W$-boson), and, as in the case of $\Gamma(h\rightarrow gg)$,
contributions from all the electrically charged s-quarks and
s-leptons through $D$-term interactions.

\begin{figure}[h!t]
\begin{center}
\includegraphics[width=4.25in]{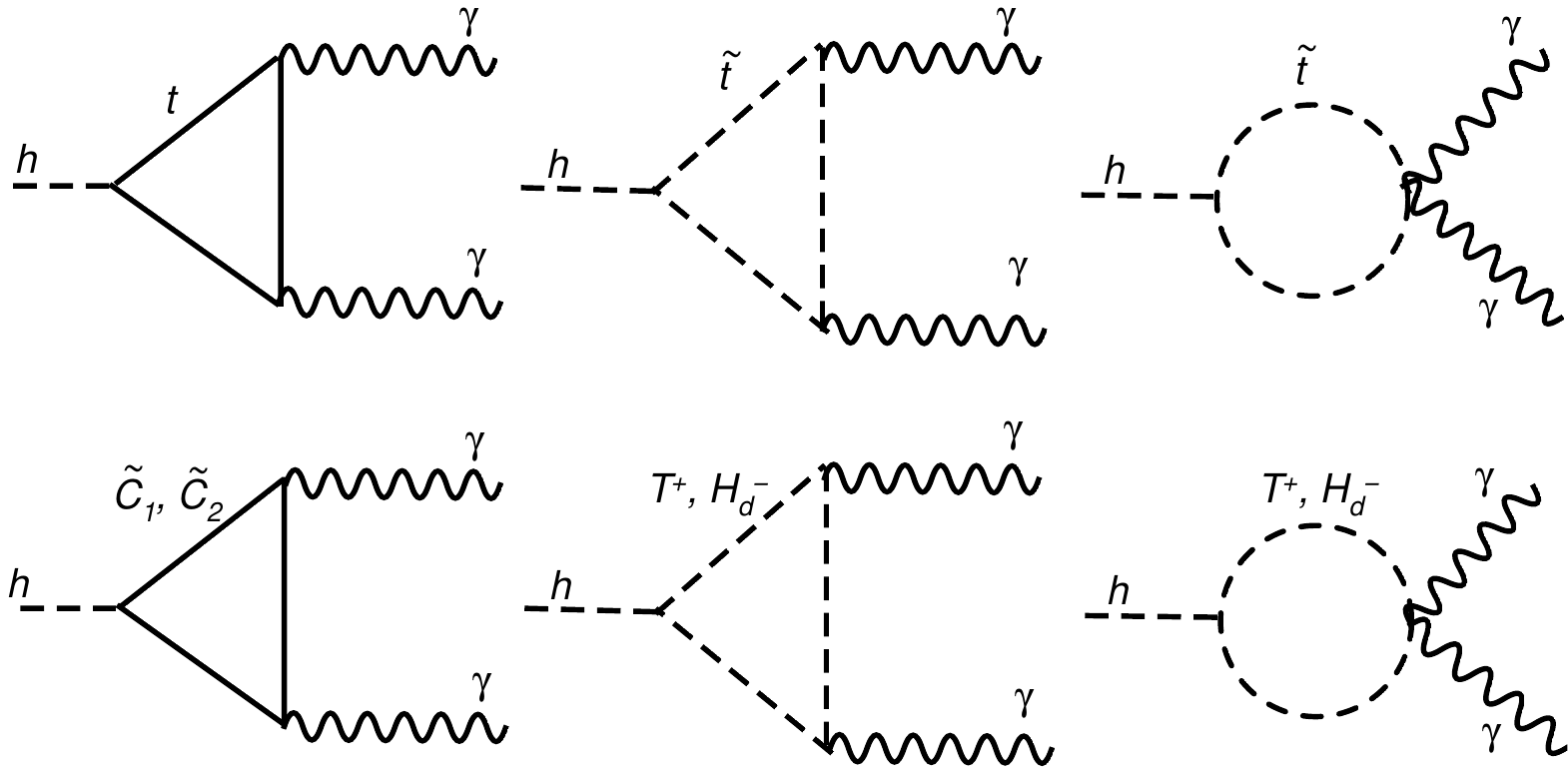}
\caption{
The diagrams that contribute to
the amplitude $\mathcal{A}(h\rightarrow\gamma\gamma)$
in the TESSM.
The top three diagrams are present in the MSSM,
and the bottom three diagrams involve the coupling
$\lambda$.
}
\label{fig:feyn-diphoton}
\end{center}
\end{figure}

In the TESSM, we have the additional contributions
from the states composed
dominantly of the charged triplets,
and also new contributions
from the MSSM matter content induced by $\lambda$ (see Fig.~\ref{fig:feyn-diphoton}).
These contributions may be important
when $\lambda$ is as large as the top Yukawa coupling,
and so in this subsection we use $\lambda=0.9$ for our numerical studies.
In this work, we will simplify our analysis by ignoring
contributions from the $D$-term interactions except those
involving the s-tops,
and, using the same approximations as in the previous subsection of
large $\tan\beta$ and $h\simeq a^{\prime}_u$, we have
the interactions and fermion masses
\begin{align}
-\mathcal{L}&\supset
2\lambda^2\,v\,h \left(\left|T^+\right|^2+ \left|H_d^-\right|^2\right)
+\lambda\,(h+v)\,
\left(\overline{\widetilde{H}^{+}}P_L \widetilde{T}^{+}
+\overline{\widetilde{T}^{+}}P_R \widetilde{H}^{+}\right)
\nonumber\\
&\quad+\mu (\overline{\widetilde{H}^{+}}\widetilde{H}^{+})
+M_T (\overline{\widetilde{T}^{+}}\widetilde{T}^{+}),
\label{eq:L-diphoton}
\end{align}
where $\widetilde{H}^{+}$ and $\widetilde{T}^{+}$
are Dirac spinors formed from the Higgsinos and
the fermionic components of the charged triplet states
\begin{align}
\widetilde{H}^{+}\equiv
\begin{pmatrix}\widetilde{H}_u^{+} \\ \widetilde{H}_d^{-\dag}\end{pmatrix},
\quad
\widetilde{T}^{+}\equiv
\begin{pmatrix}\widetilde{T}^{+} \\ \widetilde{T}^{-\dag}\end{pmatrix},
\end{align}
and $P_{L,R}$ are the projection operators
\begin{align}
P_L\equiv \begin{pmatrix}1 & 0 \\ 0 & 0\end{pmatrix},
\quad
P_R\equiv \begin{pmatrix}0 & 0 \\ 0 & 1\end{pmatrix}.
\end{align}

Although none of the charged states in Eq.~\ref{eq:L-diphoton}
is a mass eigenstate, we approximate the scalar states
as mass eigenstates with masses
\begin{align}
m^2_{T^{+}} &\simeq M_T^2+m_T^2,\nonumber\\
m^2_{H_d^{-}} &\simeq M_A^2,
\end{align}
so that the contributions of these states
to the amplitude $\mathcal{A}(h\rightarrow\gamma\gamma)$ have
the same form.
In the fermionic sector, the
contribution to the amplitude $\mathcal{A}(h\rightarrow\gamma\gamma)$
comes exclusively from the mixing between $\widetilde{H}^{+}$ and $\widetilde{T}^{+}$.
We can diagonalize the fermionic mass matrix with two unitary
transformations
\begin{align}
\begin{pmatrix}
\overline{\widetilde{H}^{+}} \\ \overline{\widetilde{T}^{+}}
\end{pmatrix}^T
V^{\dag}V
\begin{pmatrix}
\mu & \lambda v \\ 0 & M_T
\end{pmatrix}
U^{\dag}U
P_L
\begin{pmatrix}
\widetilde{H}^{+} \\ \widetilde{T}^{+}
\end{pmatrix}
=
\begin{pmatrix}
\overline{\widetilde{C}_1} \\ \overline{\widetilde{C}_2}
\end{pmatrix}^T
\begin{pmatrix}
m_{\widetilde{C}_1} & 0 \\ 0 & m_{\widetilde{C}_2}
\end{pmatrix}
P_L
\begin{pmatrix}
\widetilde{C}_1 \\ \widetilde{C}_2
\end{pmatrix},
\end{align}
where $U$ and $V$ are respectively parameterized
by $\varphi$ and $\varphi^{\prime}$,
\begin{align}
U\equiv
\begin{pmatrix} c_{\varphi} & -s_{\varphi}
\\ s_{\varphi} & c_{\varphi} \end{pmatrix},
\quad
V\equiv
\begin{pmatrix} c_{\varphi^{\prime}} & -s_{\varphi^{\prime}} \\
s_{\varphi^{\prime}} & c_{\varphi^{\prime}} \end{pmatrix},
\end{align}
where $c_{\varphi}=\cos\varphi$ and $s_{\varphi}=\sin\varphi$,
and $c_{\varphi^{\prime}}$ and $s_{\varphi^{\prime}}$ are similarly defined.
These mixing angles
are given by
\begin{align}
\tan 2\varphi&=\frac{2\lambda\,\mu\,v}{M_T^2-\mu^2+\lambda^2 v^2},\nonumber\\
\tan 2\varphi^{\prime}&=\frac{2\lambda\, v\,M_T}{M_T^2-\mu^2-\lambda^2 v^2}.
\end{align}
In terms of the mass eigenstates and mixing angles,
the chargino interactions in Eq.~\ref{eq:L-diphoton}
take the form
\begin{align}
-\mathcal{L}\supset
\lambda\,h\,
\left(
-c_{\varphi^{\prime}}s_{\varphi}
\overline{\widetilde{C}_1}\widetilde{C}_1
+c_{\varphi}s_{\varphi^{\prime}}
\overline{\widetilde{C}_2}\widetilde{C}_2
\right)+
\lambda\,h\,
\left(
c_{\varphi^{\prime}}c_{\varphi}
\overline{\widetilde{C}_1}P_L\widetilde{C}_2
-
s_{\varphi}s_{\varphi^{\prime}}
\overline{\widetilde{C}_2}P_L\widetilde{C}_1
+
\mbox{h.c.}
\right).
\end{align}

\begin{figure}[h!t]
\begin{center}
\includegraphics[width=4.25in]{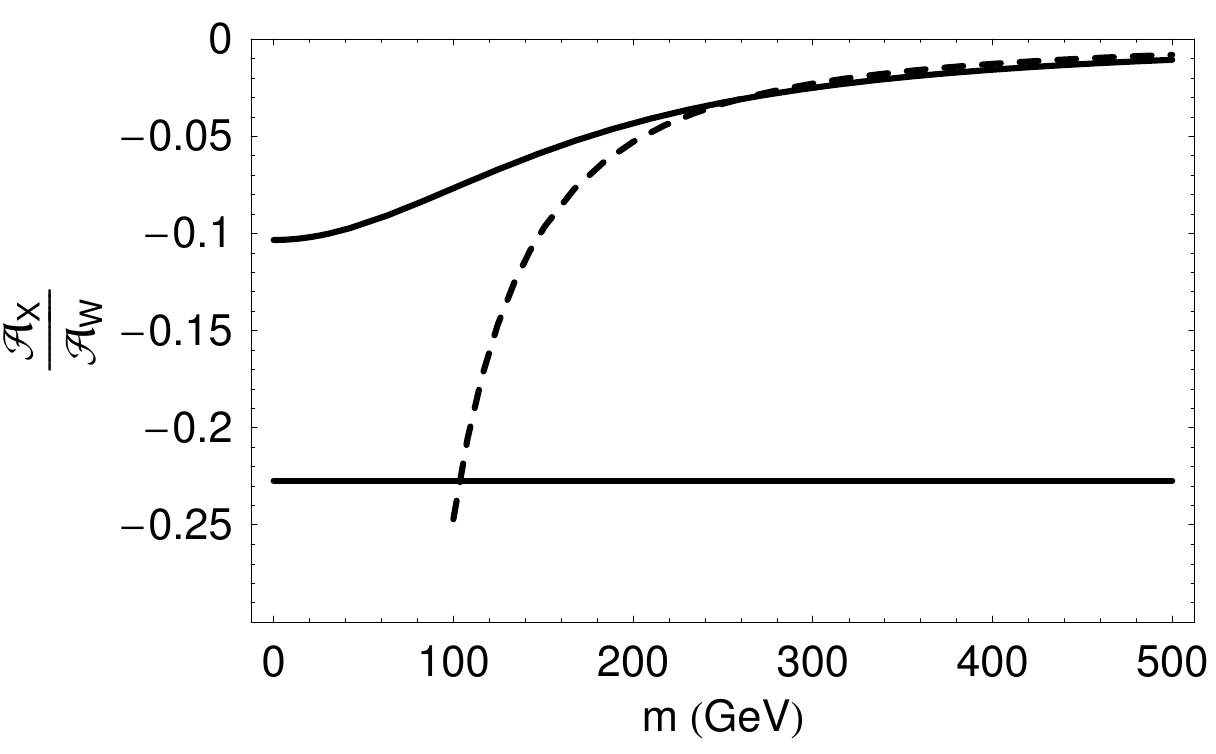}
\caption{
The ratio of amplitudes $\mathcal{A}(h\rightarrow \gamma\gamma)$
through scalar s-tops, $H_d^-$, and $T^+$ loops, compared
to the dominant $W$-boson loop contribution,
as a function, respectively, of a common
soft mass for the s-top (solid curve),
and of the mass of the states $H_d^-$ and $T^+$ (dashed line).
We use $\lambda=0.9$.
The constant, solid line denotes the top-quark contribution.
}
\label{fig:diphoton1}
\end{center}
\end{figure}

\begin{figure}[h!t]
\begin{center}
\includegraphics[width=4.25in]{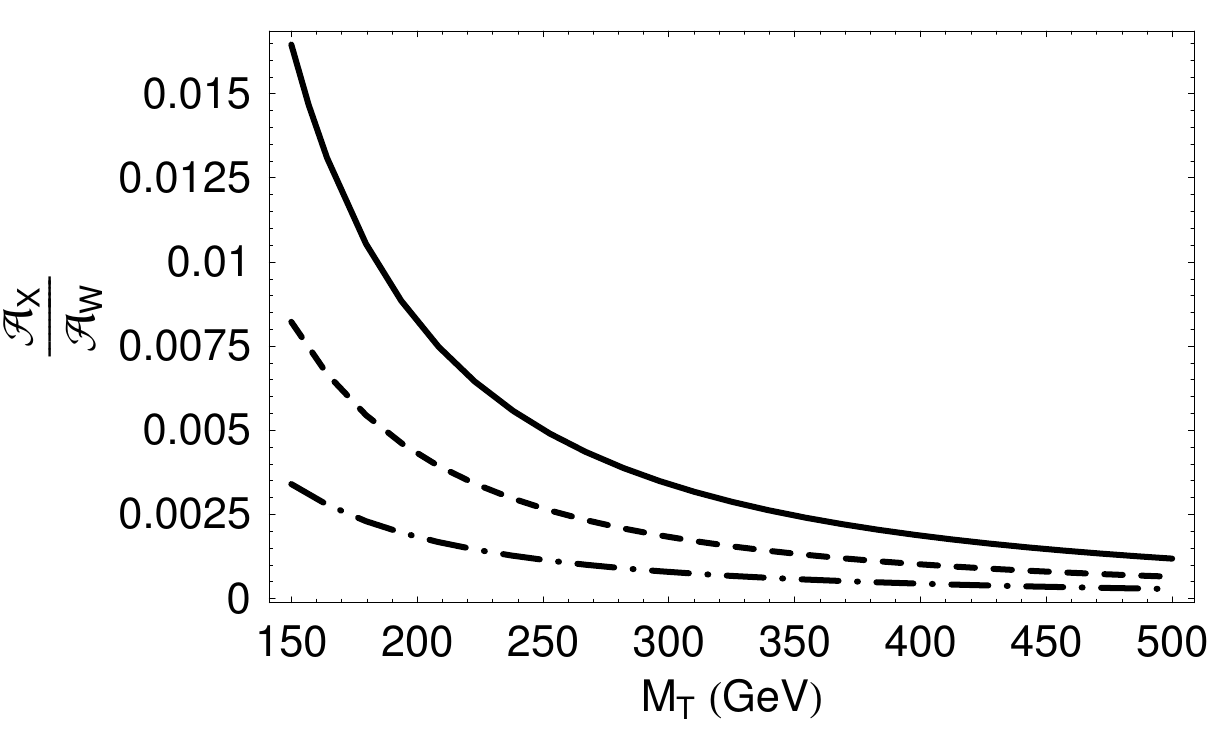}
\caption{
The sum of amplitudes $\mathcal{A}(h\rightarrow \gamma\gamma)$
given by $\widetilde{C}_1$ and $\widetilde{C}_2$ loops
normalized to the dominant $W$-boson loop amplitude,
as a function of $M_T$ for $\mu=150$ GeV (solid), 200 GeV (dashed), and 300 GeV (dot-dashed).
Note that these fermionic contributions are small
compared to the s-top contributions shown in Fig.~\ref{fig:diphoton1}.
}
\label{fig:diphoton2}
\end{center}
\end{figure}

In Figs.~\ref{fig:diphoton1} and \ref{fig:diphoton2},
we illustrate the
contributions of light s-tops and the additional
charged states to the diphoton partial decay width,
normalized with respect to the dominant $W$-boson
contribution, assuming $m_h=114$ GeV.
In Fig.~\ref{fig:diphoton1}, we show the contributions from the
top-quark (constant line), s-tops (solid line), and the charged
scalar states (dotted line).
For the s-tops (charged scalar states $H_d^-$ and $T^+$), the horizontal axis
should be interpreted as a common soft SUSY-breaking mass $M_{\smbox{SUSY}}$
(mass of these charged scalar states).
In Fig.~\ref{fig:diphoton2}, we show the sum of the fermion contributions
as a function of $M_T$ for different values of $\mu$,
and see that even for small values of $\mu$ and $M_T$
($\mu,M_T \lesssim 200$ GeV), these contributions tend to
be small.
We can partially attribute the smallness to a
cancellation between the contributions from
the two states $\widetilde{C}_1$ and $\widetilde{C}_2$,
as evident
in the relative sign
difference between the coefficients of the
$h\overline{\widetilde{C}_1}\widetilde{C}_1$ and
$h\overline{\widetilde{C}_2}\widetilde{C}_2$ interactions.
Though these fermionic contributions are small, it is interesting
to note that, while the top-quark contribution interferes
destructively with the $W$-boson contribution, the sum of these
fermionic contributions interferes constructively.
In any case, the additional $\lambda$-induced
contributions (both bosonic and fermionic) to the partial decay width $\Gamma(h\rightarrow
\gamma\gamma)$ are small compared to the s-top contributions.


Combining all these contributions to $\mathcal{A}(h\rightarrow \gamma\gamma)$,
the diphoton partial width can be significantly reduced (mostly from the s-top
contributions).
For example, with $M^2_{\smbox{SUSY}}=m_T^2=0$, $\mu=150$ GeV, $M_A=200$ GeV, and $M_T=500$ GeV,
the amplitude $\mathcal{A}(h\rightarrow\gamma\gamma)$ is decreased by
(relative to the SM) a factor of
\begin{align}
\frac{\mathcal{A}_{W}+\mathcal{A}_{t}+\mathcal{A}_{\tilde{t}}+\mathcal{A}_{H_d^-}+\mathcal{A}_{T^+}+
(\mathcal{A}_{\widetilde{C}_1}+\mathcal{A}_{\widetilde{C}_2})}
{\mathcal{A}_{W}+\mathcal{A}_{t}}
\sim \frac{1-0.23-0.11-0.05-0.008+0.001}{1-0.23}\sim 0.78,
\nonumber
\end{align}
and the diphoton decay partial width is decreased,
relative to the SM partial decay width, by a factor of
$(0.78)^2\simeq 0.6$.
We therefore can have a photo-phobic Higgs boson
in the TESSM from the contribution of
light s-tops.

\section{Fine-Tunings in TESSM}
\label{sec:FineTuning}
\subsection{Electroweak Sector}
Before discussing the fine-tuning in the electroweak sector of
the TESSM, we briefly review the little hierarchy problem
in the MSSM.
In the MSSM with large $\tan\beta$, the Higgs doublet $H_u$ is
responsible for most of the EWSB since $v\simeq\sqrt{2}\vev{H_u}$,
and it has the potential
\begin{align}
V_{H_u}=(m_{H_u}^2+\mu^2)|H_u|^2+\frac{1}{8}(g^2_2+g_1^2)|H_u|^4.
\end{align}
Minimizing the potential then gives
\begin{align}
2\vev{H_u^2}=v_{u}^2=-8\frac{m_{H_u}^2+\mu^2}{g^2_2+g_1^2},
\label{eq:MHU-vev}
\end{align}
so that
\begin{align}
m_{H_u}^2=-\frac{1}{8}(g^2_2+g_1^2)v_u^2-\mu^2.
\end{align}

Under radiative corrections, $m_{H_u}^2$ receives
large
logarithmic corrections from the s-top sector,
and we can use the renormalization group equations
to infer the value of $m_{H_u}^2$ at a fundamental
scale $\Lambda$,
\begin{eqnarray}
m_{H_u}^2(\Lambda)\simeq m_{H_u}^2(M_Z)+\frac{3 y_t^2}{8\pi^2}\left(
m^2_{\widetilde{Q}_3}+m^2_{\widetilde{\overline{U}}_3}+A_t^2
\right)
\left(\ln\frac{\Lambda}{M_Z}\right),
\end{eqnarray}
where $m^2_{\widetilde{Q}_3}$ and $m^2_{\widetilde{\overline{U}}_3}$
are the SUSY-breaking s-top masses, $y_tA_t$ is
the coupling of the trilinear interaction
$\widetilde{Q}_3H_u\widetilde{\overline{U}}_3$, and $\Lambda$ can
be taken as the scale of SUSY-breaking.
The large radiative correction leads to fine-tuning $f_s$
because the electroweak scale $v$ depends sensitively
on the value of $m^2_{H_u}$ at the fundamental scale of SUSY-breaking $\Lambda$.
We can quantify this fine-tuning as \cite{Barbieri:1987fn}
\begin{align}
f_s\equiv\frac{\delta \ln v^2}{\delta \ln m_{H_u}^2(\Lambda)}
\simeq
\frac{3 y_t^2}{4\pi^2}\left(\frac{
m^2_{\widetilde{Q}_3}+m^2_{\widetilde{\overline{U}}_3}+A_t^2}
{M_Z^2}
\right)
\left(\ln\frac{\Lambda}{M_Z}\right).
\label{eq:HiggsFineTune0}
\end{align}
As a reference of comparison,
for $m^2_{\widetilde{Q}_3}=m^2_{\widetilde{\overline{U}}_3}=A_t=1$ TeV,
and $\Lambda=10^3$ TeV, we have $f_s=80$ so that
the Higgs sector needs to be fine-tuned to one part in 80.
Thus, even though
the electroweak scale is no longer
quadratically sensitive to the fundamental scale $\Lambda$
with softly-broken SUSY,
it is quadratic sensitive to the s-top masses and trilinear
coupling $A_t$, which are required to be large
to have a Higgs mass that satisfies the LEP bounds,
and this leads to a fine-tuning in the Higgs sector
of about one part in 100.  This is the little-hierarchy
problem in the MSSM.

We can also define other measures of
fine-tuning when given a more fundamental
theory (for example, an organizing
principle of the soft SUSY-breaking
parameters) \cite{Anderson:1994dz}\cite{Anderson:1994tr}\cite{Athron:2007ry} .
However, in this work we are mainly interested in
the low-energy phenomenology of the TESSM without
appealing to a particular fundamental theory,
and we will simply define fine-tuning as in Eq.~\ref{eq:HiggsFineTune0}.

\begin{figure}[h!t]
\begin{center}
\includegraphics[width=\picwidth]{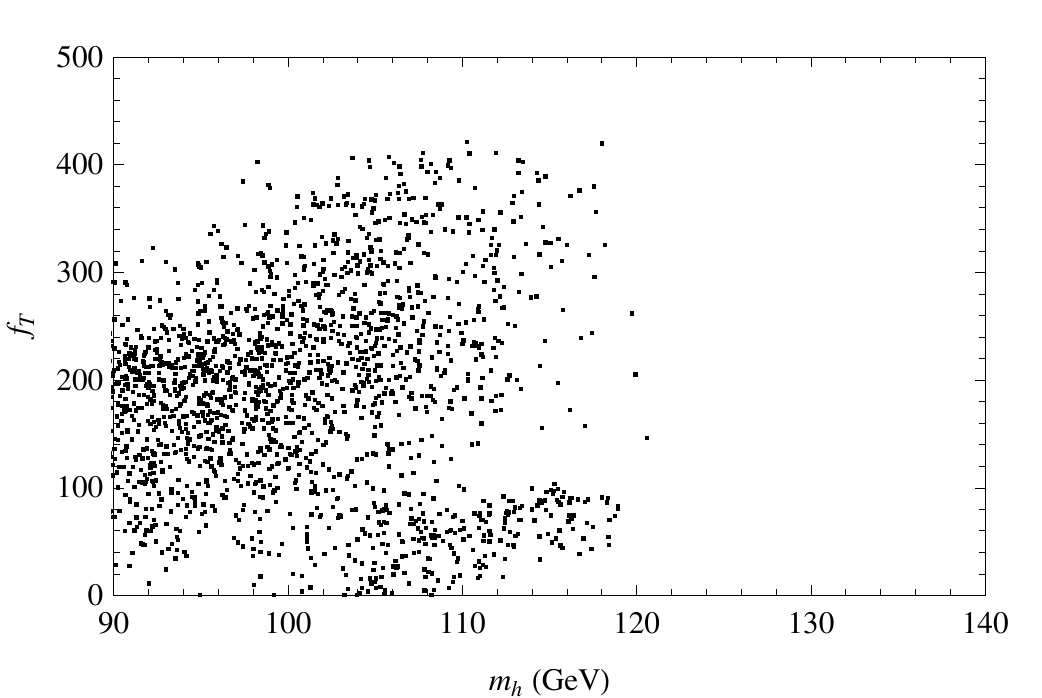}
\includegraphics[width=\picwidth]{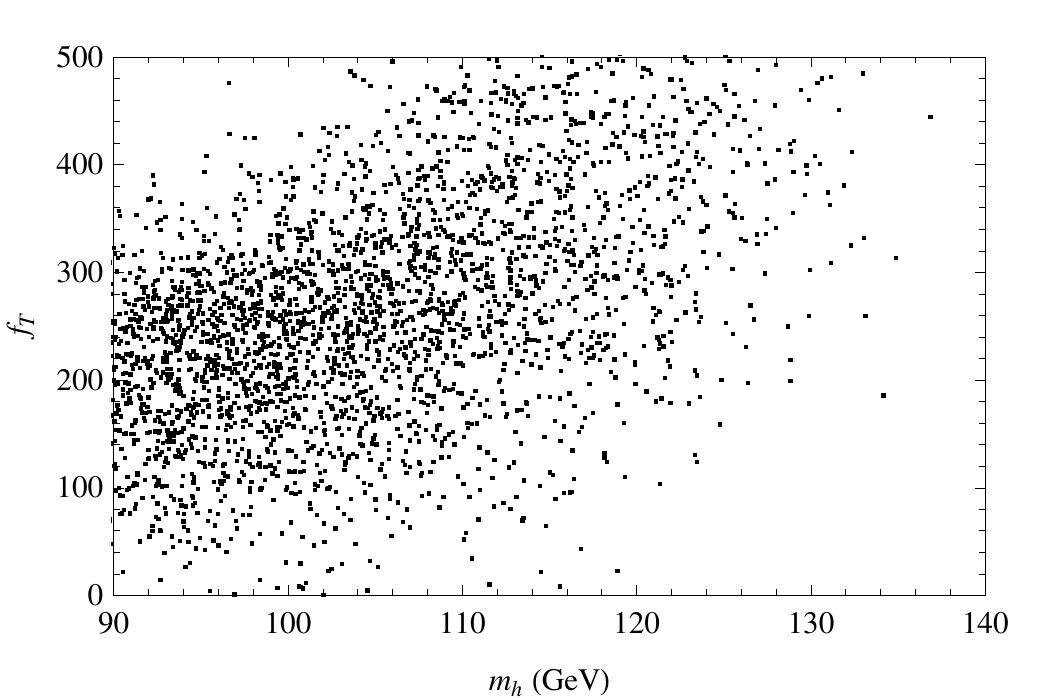}
\caption{
Fine-tuning (as defined in Eq.~\ref{eq:HiggsFineTune})
as a function of the mass lightest $CP$-even Higgs boson.
This is typically less than the fine-tuning of the MSSM
(as defined in Eq.~\ref{eq:HiggsFineTune0}) and
the NMSSM.
The plot on the left has $\lambda = 0.8$, and
the plot on the right has $\lambda = 0.9$.
}
\label{fig:NTunePlot}
\end{center}
\end{figure}

In the TESSM with $\lambda$ comparable to the top Yukawa coupling,
we do not need heavy s-top masses nor
significant mixing in the s-top sector for the Higgs
mass to satisfy the LEP bound, and as such
there is little or no fine-tuning from the s-top sector.
On the other hand, $m_{H_u}^2$ now receives radiative
corrections from the triplet sector as well as the s-top sector
\begin{eqnarray}
m_{H_u}^2(\Lambda)\simeq m_{H_u}^2(M_Z)+\frac{3 y_t^2}{8\pi^2}\left(
m^2_{\widetilde{Q}_3}+m^2_{\widetilde{\overline{U}}_3}+A_t^2
\right)
\left(\ln\frac{\Lambda}{M_Z}\right)
+
\frac{3\lambda^2}{8\pi^2}\left(
m^2_{T}+A_{\lambda}^2
\right)
\left(\ln\frac{\Lambda}{M_Z}\right),
\end{eqnarray}
and we can follow the same steps and reasoning as before to
have an estimate of
the fine-tuning due to the triplet sector $f_T$
\begin{align}
f_T
\simeq
\frac{3 \lambda^2}{4\pi^2}\left(\frac{
m^2_T+A_{\lambda}^2}
{M_Z^2}
\right)
\left(\ln\frac{\Lambda}{M_Z}\right),
\label{eq:HiggsFineTune}
\end{align}
so that $f_T=40$, for example, would mean
a tuning in $m^2_{H_u}(\Lambda)$ to one part
in 40.
The value of $f_T$ indicates
the percent change in $v^2$
per a one-percent change in $m^2_{H_u}$
at a fundamental scale of SUSY-breaking,
Generally, with large $\lambda$, for a given mass of the lightest
$CP$-even Higgs boson, the fine-tuning in $m_{H_u}^2$ is less in
the TESSM than the MSSM.
In Fig.~\ref{fig:NTunePlot}, we plot $f_T$ for
the data points shown in Fig.~\ref{fig:NLoopPlot},
where we see a rough general trend of increasing fine-tuning
with increasing Higgs mass.
On the other hand, it is possible
to have points with relatively small $f_T$ $(f_T\lesssim 20)$ that satisfy the LEP2 bound of $m_h > 114.4$ GeV,
as demonstrated in Point 1 of Table \ref{tb:SamplePoints}.
This is a great improvement over the MSSM,
and it is a consequence of the large tree-level mass
we can obtain in TESSM, so we do not have to rely on large radiative
corrections from $m_T^2$ and $A_{\lambda}$.

\subsection{Triplet Sector}
The vev of $T^0$ is induced by the vev's of the Higgs
doublets because the vev's of the Higgs doublets  $v_{u,d}$ induce a
tadpole from the trilinear interactions of the form $HTH$
in the second line of Eq.~\ref{eq:VMIX}.
We noted earlier that some cancellation
between \emph{a priori}
unrelated parameters
($\mu$ and $M_T\sin 2\beta$, for example) is required
to keep $v_t$ (and thus $\Delta T$) small
and this leads to fine-tuning in the triplet sector.
However, it is worth pointing out that $v_t$
here does not receive a large radiative
correction that requires a fine-tuning as severe as
the fine-tuning in the hierarchy problem
in the triplet-extended standard model potential
analyzed in Chivukula et al.~\cite{SekharChivukula:2007gi}.
It is easiest to see this in the limit $m_T^2=B_T=A_{\lambda}^2=0$
(SUSY-limit in the triplet sector) where
the triplet vev $v_t$
in Eq.~\ref{eq:vtCond} takes a particularly simple form
\begin{align}
v_t &= \frac{\sqrt{2}}{2}
(\lambda v^2)\
\frac{\mu-M_Ts_{\beta}c_{\beta}}
{M_T^2+\tfrac{\lambda^2}{2}v^2},
\label{eq:vtCondSimp}
\end{align}
and the 1-loop corrections to $v_t$ then
involve 1-loop corrections to the parameters
$\lambda$, $v_{u,d}$, and $M_T$.
The parameters $\lambda$, $\mu$, and $M_T$
come from the superpotential,
and the nonrenormalization theorem
dictates that the radiative corrections
to these parameters run only in a
logarithmical manner due to
wavefunction renormalizations only.
Though the loop corrections to $v_{u,d}$
may require
a fine-tuning of one part in a few hundreds
(this is the little hierarchy
problem in the MSSM), this is much more benign than
the fine-tuning in the triplet-extended SM studied
in Chivukula et al.~\cite{SekharChivukula:2007gi}.

On the other hand, there is a source of fine-tuning
in $v_t$ because we often require some degree of
cancellation to make $\Delta T$
small.
We can define a quantitative measure of fine-tuning
in $\Delta T$ by
\begin{align}
\kappa_T&\equiv\frac{\delta\ln \Delta T}{\delta\ln M_T}
=2\frac{\delta\ln v_t}{\delta\ln M_T}\nonumber\\
&=
\left(\frac{2M_T}{\sin 2\beta(A_{\lambda}+M_T)-2\mu}\right)
\left(\frac{4\mu M_T+\sin 2\beta(m_T^2+B_T-2A_{\lambda}M_T-M_T^2+\frac{\lambda^2}{2}v^2)}
{M_T^2+m_T^2+B_T+\frac{\lambda^2}{2}v^2}\right),
\label{eq:TripletFineTune}
\end{align}
so that $\kappa_T$ is large when there is a large
cancellation in the combination
\begin{align}
\sin 2\beta(A_{\lambda}+M_T)-2\mu,
\nonumber
\end{align}
that makes $\Delta T$ unnaturally small.

The definition in Eq.~\ref{eq:TripletFineTune},
however, may not be satisfactory
because it does not take into account
the range of allowed $\Delta T$.
For example, for the parameters listed in
Eq.~\ref{eq:Texample}
\begin{align}
\tan\beta=3,\quad\lambda =0.9,\quad \mu =150\ \mbox{GeV},\nonumber\\
m_T^2=B_T=A_{\lambda}^2= (200\ \mbox{GeV})^2,
\nonumber
\end{align}
we have viable $\Delta T$ in the regions
\begin{align}
250\ \mbox{GeV} < M_T < 375\ \mbox{GeV},\
\mbox{or}\ M_T> 3.0\ \mbox{TeV},
\nonumber
\end{align}
and it may be reasonable to expect that
any value of $M_T$ in the small range between 250 GeV
and 375 GeV is equally fine-tuned.
However, Eq.~\ref{eq:TripletFineTune} would
give different values of $\kappa_T$ for different values of $M_T$,
and may even diverge if
$M_T$ is such that we have $v_t=0$.
It is true that when $v_t=0$ we have unnatural, complete
cancellation, but in our work we only use $v_t$ in a binary way:
to distinguish cases with viable $\Delta T$ from those with unacceptably
large $\Delta T$.
Once $v_t$ is small enough
to have viable $\Delta T$, we do not care whether
$v_t=1$ GeV or $v_t=0.01$ GeV, for example.

As in Section \ref{sec:TESSM}, we can also estimate the
fine-tuning in $\Delta T$ due to $M_T$ as shown in Athron et
al.~\cite{Athron:2007ry} when there is a cancellation in the
numerator of Eq.~\ref{eq:ObliqueTFormula} that makes $\Delta T$
small.
With all parameters other than $M_T$ fixed,
we first compute $M_T^{\ast}$ such
that for $M_T> M_T^{\ast}$, $\Delta T$ is always
viable ($\Delta T<0.1$), and define fine-tuning as
\begin{align}
\kappa^{\prime}\equiv
\frac{M_T^{\ast}}{\mbox{Range of}\ M_T(\mbox{with}\ M_T<M_T^{\ast})\ \mbox{that gives
viable}\ \Delta T}.
\label{eq:TripletFineTune2}
\end{align}
This definition of
fine-tuning is harder to implement because,
given a set of parameters except $M_T$,
we first have to find out if
regions of $M_T$ allowed by $\Delta T$
comes about because of cancellations,
before we can apply Eq.~\ref{eq:TripletFineTune2}.
For example,
it is possible that $\Delta T$ is always
viable for any value of $M_T$ (as are the cases
for Points 3 through 6 of Table~\ref{tb:SamplePoints}),
so that we can not apply Eq.~\ref{eq:TripletFineTune2}
as there is no fine-tuning in $\Delta T$.
Despite its limited applicability
compared to $\kappa_T$,
$\kappa^{\prime}_T$ may be a more reasonable
measure of fine-tuning when there is a cancellation
that leads to a small value for $\Delta T$.
For Point 1(2)  in Table~\ref{tb:SamplePoints},
we have $\kappa_T\sim 33(11)$ and $\kappa^{\prime}_T\sim 4.7(6.4)$,
corresponding to a 33(11)\% change in $\Delta T$ per a 1\% change in $M_T$,
and also cancellation of one part in 4.7(6.4).
For the other four points in Table~\ref{tb:SamplePoints}
where $\kappa^{\prime}_T$ in Eq.~\ref{eq:TripletFineTune2}
is not well-defined,
the values of $\kappa_T$ are small, indicating small
fine-tuning for these sets of parameters.
Since a complete analysis of fine-tuning in the triplet sector in
the TESSM is outside the scope of this work, we will conclude this
section noting that in an extreme case (Eq.~\ref{eq:Texample}),
$\kappa^{\prime}_T\sim 24$, so we suspect that the typical
fine-tuning in the triplet sector be less than one part in 24.


\section{Conclusions}
\label{sec:Conclusion}
In this work we have revisited a very simple extension
to the MSSM by adding a hypercharge-neutral,
$SU(2)$ triplet chiral superfield.
We considered this model as a reasonably economical extension of the MSSM and an alternative to the NMSSM,
and extended
the phenomenological studies in several directions.
In addition to discussing the decoupling behavior of the triplets
and comparing it to the decoupling behavior of the singlet of the NMSSM,
we have computed the mass of the
lightest $CP$-even Higgs boson to
one-loop in the large quartic coupling $\lambda$.
With $\lambda$, the Higgs-triplet-Higgs coupling in the
superpotential, being comparable with the top Yukawa coupling, we find that
the model is able to satisfy LEP2
bounds on the Higgs mass without contributions
from the s-top sector.
At the expense of perturbativity at the GUT scale,
we have checked that the model can give much smaller
fine-tuning in the
electroweak sector than the MSSM.
In the triplet sector, there may be fine-tuning
in having small oblique corrections, but
we estimate this fine-tuning to be no worse than
about one part in 30.

With large $\lambda$, the TESSM opens up previously forbidden
regions of parameters in the MSSM.
In particular, both s-tops can be light in
the TESSM.
The light s-tops can then lead to phenomenology
that is very different from the MSSM with
important implications for the LHC, such
as a Higgs boson that is both gluo-philic and
photo-phobic.

Our simple analysis here can be extended
in many directions, and these further studies
must be done if the model is going to make
precise predictions at the LHC.
With large $\lambda$, there can be important
higher-loop effects to the mass of the lightest, $CP$-even
Higgs boson.
Furthermore, important higher-loop QCD effects
must also be included to properly study the gluon-gluon fusion
production and the diphoton decay of the Higgs boson.
We leave these open projects for the future and
hope they may add to the already-rich possibilities
of phenomenology that will be seen at the LHC.

\appendix
\section{Field-dependent mass matrices}
\label{app:Field-dep-masses}
In this appendix, we list the field-dependent matrices that enter into
the Coleman-Weinberg potential in Eq.~\ref{eq:CWex01}.
We have five mass matrices, one for each set of particles:
the $CP$-even Higgs bosons ($\mathcal{M}_a$),
the $CP$-odd Higgs bosons ($\mathcal{M}_b$),
the charged Higgs bosons ($\mathcal{M}_c$),
the neutralinos ($\mathcal{M}_{\widetilde{N}}$),
and the charginos ($\mathcal{M}_{\widetilde{C}}$).
We first list the elements of the Higgs bosons.
\begin{align}
\left(\mathcal{M}^2_{a}\right)_{11} &= m_{H_u }^2 +\mu^2 + \frac{1}
{8}\left( {g_1^2  + g_2^2 } \right)\left( {3\,a_u^2  - a_d^2 } \right)\,+ \frac{{\lambda ^2 }}
{2}\,\left( {a_t^2  + a_d^2 } \right)- \sqrt 2 \,\lambda \,\mu \,a_t ,
\hfill \\
\left(\mathcal{M}^2_{a}\right)_{12} &=  - B_\mu -
\frac{1}{4}\left( g_1^2  + g_2^2 \right)\,a_u \,a_d  +\lambda^2\, a_u\, a_d
+ \frac{\lambda }{{\sqrt 2 }}\,\left( {A_\lambda   + M_T } \right)\,a_t , \hfill \\
\left(\mathcal{M}^2_{a}\right)_{13} &=   \lambda^2\,a_u \,a_t  - \sqrt 2\lambda \,\mu\,a_u
+ \frac{\lambda}{\sqrt{2}}\left( {A_\lambda   + M_T } \right)\,a_d, \hfill \\
\left(\mathcal{M}^2_{a}\right)_{22} &= m_{H_d }^2 +\mu ^2 + \frac{{\lambda ^2 }}
{2}\,\left( {a_t^2  + a_u^2 } \right) + \frac{1}
{8}\left( {g_1^2  + g_2^2 } \right)\,\left( {3\,a_d^2  - a_u^2 } \right)- \sqrt 2 \,\lambda \,\mu \,a_t, \hfill \\
\left(\mathcal{M}^2_{a}\right)_{23} &=   \lambda^2\,a_d \,a_t  - \sqrt 2\lambda \,\mu\,a_d
+ \frac{\lambda}{\sqrt{2}}\left( {A_\lambda   + M_T } \right)\,a_u, \hfill \\
\left(\mathcal{M}^2_{a}\right)_{33} &= M_T^2  + m_T^2  + B_T  + \frac{{\lambda ^2 }}
{2}\,\left( {a_d^2  + a_u^2 } \right), \hfill
\end{align}

\begin{align}
\left(\mathcal{M}^2_{b}\right)_{11} &= m_{H_u }^2 +\mu^2 + \frac{1}
{8}\left( {g_1^2  + g_2^2 } \right)\,\left( {a_u^2  - a_d^2 } \right)
+ \frac{{\lambda ^2 }}{2}\,\left( {a_t^2  + a_d^2 } \right) - \sqrt 2 \,\lambda \,\mu \,a_t  , \hfill \\
\left(\mathcal{M}^2_{b}\right)_{12} &= B_\mu   - \frac{\lambda }
{{\sqrt 2 }}\,\left( {M_T   +  A_\lambda} \right)\,a_t , \hfill \\
\left(\mathcal{M}^2_{b}\right)_{13} &= \frac{\lambda }
{{\sqrt 2 }}\,\left( {M_T  - A_\lambda  } \right)\,a_d , \hfill \\
\left(\mathcal{M}^2_{b}\right)_{22} &=   m_{H_d }^2 +\mu^2+ \frac{1}
{8}\left( {a_d^2  - a_u^2 } \right)\,\left( {g_1^2  + g_2^2 } \right)
+ \frac{{\lambda ^2 }}{2}\,\left( {a_t^2  + a_u^2 } \right)- \sqrt 2 \,\lambda \,\mu \,a_t , \hfill \\
\left(\mathcal{M}^2_{b}\right)_{23} &= \frac{\lambda }
{{\sqrt 2 }}\,\left( {M_T  - A_\lambda  } \right)\,a_u , \hfill \\
\left(\mathcal{M}^2_{b}\right)_{33} &= M_T^2  + m_T^2  - B_T  + \frac{{\lambda ^2 }}
{2}\,\left( {a_d^2  + a_u^2 } \right), \hfill
\end{align}

\begin{align}
  \left(\mathcal{M}^2_{c}\right)_{11} &= m_{H_u }^2 +\mu ^2 + \left( {\lambda ^2  - \frac{{g_1^2  - g_2^2 }}
{8}} \right)\,a_d^2 +\frac{1}{8}\,\left(g_1^2  + g_2^2\right)\,a_u^2  +
\sqrt 2\lambda \,\mu\,a_t  + \frac{\lambda^2 }{2}\,a_t^2, \hfill \\
\left(\mathcal{M}^2_{c}\right)_{12} &= B_\mu   + \frac{1}
{2}\,\left( {\lambda ^2  + \frac{{g_2^2 }}
{2}} \right)\,{a_d \,a_u } + \frac{{\lambda}}
{{\sqrt 2 }}\,\left( {M_T  + A_\lambda  } \right)\,a_t, \hfill \\
  \left(\mathcal{M}^2_{c}\right)_{13} &= {\lambda \,\mu\,a_u  + \frac{{1}}
{{\sqrt 2 }}\,\left( {\lambda ^2  - \frac{{g_2^2 }}
{2}} \right)}\,a_u\,a_t - \lambda\,M_T\,a_d , \hfill \\
  \left(\mathcal{M}^2_{c}\right)_{14} &= {\lambda \,\mu\,a_u  - \frac{{1 }}
{{\sqrt 2 }}\,\left( {\lambda ^2  - \frac{{g_2^2 }}
{2}} \right)} \,a_u\,a_t - \lambda\,A_\lambda\,a_d  , \hfill \\
  \left(\mathcal{M}^2_{c}\right)_{22} &= m_{H_d }^2 +\mu ^2 + \left( {\lambda ^2  - \frac{{g_1^2  - g_2^2 }}
{8}} \right)\,a_u^2 +\frac{1}{8}\,\left(g_1^2  + g_2^2\right)\,a_d^2  +
\sqrt 2\lambda\,\mu\,a_t  + \frac{\lambda^2 }{2}\,a_t^2, \hfill \\
  \left(\mathcal{M}^2_{c}\right)_{23} &=  - {\lambda \,\mu\,a_d  + \frac{{1}}
{{\sqrt 2 }}\,\left( {\lambda ^2  - \frac{{g_2^2 }}
{2}} \right)} \,a_d\,a_t + \lambda\,A_\lambda\,a_u  , \hfill \\
\left(\mathcal{M}^2_{c}\right)_{24} &=  - {\lambda \,\mu\,a_d - \frac{{1}}
{{\sqrt 2 }}\,\left( {\lambda ^2  - \frac{{g_2^2 }}
{2}} \right)}\,a_d\,a_t + \lambda\,M_T \,a_u  , \hfill \\
\left(\mathcal{M}^2_{c}\right)_{33} &= M_T^2  + m_T^2  + \frac{{g_2^2 }}
{4}\,\left( {a_d^2  + 2\,a_t^2  - a_u^2 } \right)+ \lambda ^2 \,a_u^2  , \hfill \\
\left(\mathcal{M}^2_{c}\right)_{34} &= B_T  - \frac{{g_2^2}}
{2}\,a_t^2 , \hfill \\
\left(\mathcal{M}^2_{c}\right)_{44} &= M_T^2  + m_T^2  + \frac{{g_2^2 }}
{4}\,\left( {a_u^2  + 2\,a_t^2  - a_d^2 } \right)+ \lambda ^2 \,a_d^2,\hfill
\end{align}
where $m_{H_{u,d}}^2$ satisfy the minimization conditions
Eqs.~\ref{eq:Min-MHU} and \ref{eq:Min-MHD}.

For the neutralino and charginos, since we do not take into account mixing with
the gauginos, we have reduced matrices compared to
those in Eqs.~\ref{eq:NMatrix} and \ref{eq:CMatrix},
and here we can simply replace the vevs by the corresponding particle
\begin{align}
\mathcal{M}_{\widetilde{N}}&=
\begin{pmatrix}
 0 & -\mu+\frac{\lambda}{\sqrt{2}}a_t & \frac{1}{\sqrt{2}}\lambda a_u \\
 -\mu+\frac{\lambda}{\sqrt{2}}a_t & 0 & \frac{1}{\sqrt{2}}\lambda a_d \\
\frac{1}{\sqrt{2}}\lambda a_u & \frac{1}{\sqrt{2}}\lambda a_d & M_T
\end{pmatrix},
\end{align}
\begin{align}
\mathcal{M}_{\widetilde{C}}&=
\begin{pmatrix}
\mu+\frac{\lambda}{\sqrt{2}}a_t & -\lambda a_d&  \\
 \lambda a_u & M_T
\end{pmatrix}.
\end{align}

\section{Diphoton decay width of a real scalar}
\label{app:di-photon-width}

In this appendix, we review the formula for the decay width of a
real scalar $\phi^0$ (with mass $m_{\phi}$) decaying into two
photons $\Gamma(\phi^0\rightarrow\gamma\gamma)$
\cite{Gunion:1984yn}.
Generally, given the interactions
\begin{align}
\mathcal{L}\supset
-A_s s^{+}s^{-}\phi^0-\frac{A_{\psi}}{2}\phi^0\overline{\psi}\psi + A_W W^{+\mu}W^{-}_{\mu}\phi^0,
\end{align}
where $s^{\pm}$ ($\psi$) $\{W_{\mu}^{\pm}\}$ is a charged
scalar (fermion) $\{$gauge boson$\}$ with
mass $m_s$ ($m_\psi$) \{$m_W$\} and electric charge $Q_s$
$(Q_\psi)$ $\{Q_W\}$,
the diphoton partial decay width is given by
\begin{align}
\Gamma(\phi^0\rightarrow\gamma\gamma)
=\frac{\alpha^2_{\smbox{em}}}{1024\pi^3}m_{\phi}
\left|
N_{\psi} A_{\psi} Q_{\psi}^2 \frac{m_{\phi}}{m_{\psi}} F_{\psi}
+
N_s A_s Q_s^2 \frac{m_{\phi}}{m^2_s} F_s
+
N_W A_W Q_W^2 \frac{m_{\phi}}{m^2_W} F_W
\right|^2,
\label{eq:Di-photon-formula}
\end{align}
where $N_i$ are factors to account for
additional degrees of freedom (such as color) and
\begin{align}
F_s&=\tau_s\left[1-\tau_s f(\tau_s)\right],\\
F_{\psi}&=-2\tau_{\psi}\left[1+(1-\tau_{\psi})f(\tau_{\psi})\right],\\
F_W&=2 + 3\tau_W +3\tau_W(2-\tau_W)f(\tau_W),
\end{align}
where
\begin{align}
\tau_i &\equiv 4\frac{m_i^2}{m_{\phi}^2},\quad\mbox{for}\ i=s,\psi, W,\\
f(\tau)&=\begin{cases}
\left(\arcsin\sqrt{\frac{1}{\tau}}\right)^{2} & \mbox{if}\ \tau>1,
\\
-\frac{1}{4}\left(\ln\frac{\eta_+}{\eta_-}-i\pi\right)^2 & \mbox{if}\ \tau<1,
\end{cases},\\
\eta_{\pm}&\equiv 1\pm \sqrt{1-\tau}.
\end{align}

In the case of colored particles, we can make the replacement
\begin{align}
NQ^4\alpha^2_{\smbox{em}}\rightarrow 2\alpha^2_s
\label{eq:diphoton-to-digluon}
\end{align}
to compute the di-gluon decay width $\Gamma(\phi^0\rightarrow gg)$,
which is related to the gluon-gluon fusion production cross section by
\begin{align}
\sigma(gg\rightarrow \phi^0)=\frac{\pi^2}{8m_{\phi}^3}\Gamma(\phi^0\rightarrow gg).
\label{eq:sigma-gamma-conv}
\end{align}

\begin{acknowledgments}
This work is supported by the US National
Science Foundation under Grants
No.~PHY-0354226, No.~PHY-0555545, and No.~PHY-0354838.
We are indebted to R.~Sekhar Chivukula and Elizabeth S.~Simmons
for inspiring this project and many useful
discussions.
We would also like to thank Neil D.~Christensen
and Puneet Batra
for helpful comments and suggestions.
The Feynman diagrams in this work are
drawn using \verb"JaxoDraw" \cite{Binosi:2003yf},
and we check some
of our results using \verb"hdecay" \cite{Djouadi:1997yw}.
\end{acknowledgments}

\end{document}